\documentclass[twocolumn,english,prd,amsmath,amssymb,superscriptaddress]{revtex4}
\usepackage{graphicx}
\usepackage{dcolumn}
\usepackage{bm}

\begin{document}

\newcommand{\lexp}{\mathop{\langle}}
\newcommand{\rexp}{\mathop{\rangle}}
\newcommand{\rexpc}{\mathop{\rangle_c}}
\newcommand{\beq}{\begin{equation}}
\newcommand{\eeq}{\end{equation}}
\newcommand{\beqa}
{\begin{eqnarray}}
\newcommand{\eeqa}{\end{eqnarray}}

\def\k{{\hbox{\bf k}}}
\def\q{{\hbox{\bf q}}}
\def\x{{\hbox{\bf x}}}
\def\r{{\hbox{\bf r}}}
\def\v{{\hbox{\bf v}}}
\def\u{{\hbox{\bf u}}}
\def\dD{\delta_{\rm D}} 
\def\tvk{{\hat{\k}}}
\def\tvq{{\hat{\q}}}

\def\Dp{D_+}
\def\de{\delta}
 
\def\Mpc{\, h^{-1} \, {\rm Mpc}}
\def\Gpc{\, h^{-1} \, {\rm Gpc}}
\def\Gpccube{\, h^{-3} \, {\rm Gpc}^3}
\def\kvecMpc{\, h \, {\rm Mpc}^{-1}}
\def\la{\mathrel{\mathpalette\fun <}}
\def\ga{\mathrel{\mathpalette\fun >}}
\def\fun#1#2{\lower3.6pt\vbox{\baselineskip0pt\lineskip.9pt
        \ialign{$\mathsurround=0pt#1\hfill##\hfil$\crcr#2\crcr\sim\crcr}}}
        
\title{Nonlinear Evolution of Baryon Acoustic Oscillations}

\author{Mart\'{\i}n  Crocce}\email{hmc238[at]nyu.edu}
\affiliation{Center for Cosmology and Particle Physics,\\
Department of Physics, New York University, New York, NY 10003 }
\affiliation{Institut de Ci\`encies de l'Espai, IEEC-CSIC, Campus UAB,
Facultat de Ci\`encies, Torre C5 par-2,  Barcelona 08193, Spain}

\author{Rom\'an Scoccimarro}\email{rs123[at]nyu.edu}
\affiliation{Center for Cosmology and Particle Physics,\\
Department of Physics, New York University, New York, NY 10003 }

\begin{abstract}

We study the nonlinear evolution of the baryon acoustic oscillations (BAO) in the dark matter power spectrum and correlation function using renormalized perturbation theory (RPT).  In a previous paper we showed that RPT successfully predicts the damping of acoustic oscillations; here we extend our calculation to the enhancement of power due to mode-coupling. We show that mode-coupling generates additional oscillations that are out of phase with those in the linear spectrum, leading to shifts in the scales of oscillation nodes defined with respect to a smooth spectrum. When Fourier transformed, these out of phase oscillations induce percent-level shifts in the acoustic peak of the two-point correlation function. We present predictions for these shifts as a function of redshift; these should be considered as a robust lower limit to the more realistic case that  includes in addition redshift distortions and galaxy bias. We show that these nonlinear effects occur at very large scales, leading to a breakdown of linear theory at scales much larger than commonly thought.  We discuss why  virialized halo profiles are not responsible for these effects, which can be understood from basic physics of gravitational instability. Our results are in excellent agreement with numerical simulations, and can be used as a starting point for modeling BAO in future observations.  To meet this end, we suggest a simple physically motivated model to correct for the shifts caused by mode-coupling.  

\end{abstract}

\maketitle

\section{Introduction}

The imprint of baryon acoustic oscillations (BAO) in the power spectrum and two-point correlation function of galaxies has emerged as a potentially robust probe of the cause of the present acceleration of the universe, through their sensitivity to the angular diameter distance and the Hubble constant as a function of redshift.  Measurements of this signature can be used to help probe dark energy or large-distance modifications of gravity as the explanation for the acceleration of the universe. 

The basic idea behind the BAO method is to use the characteristic size of the sound horizon imprinted in the spatial correlation properties of galaxies or other tracers as a standard ruler~\cite{BAO}. In the power spectrum, this characteristic scale and its harmonics lead to oscillatory features, the so-called ``baryonic wiggles"~\cite{wiggles}. In the two-point correlation function, these features translate to a broad peak at the sound horizon scale, about $100 \Mpc$ for acceptable cosmological parameters. The BAO have been detected recently in both the correlation function and power spectrum of galaxies~\cite{BAOdetect,BAOobs}.

This acoustic signature evolves with time, and the key issue is how. According to linear perturbation theory the acoustic signature increases in amplitude but its spatial pattern remains static, i.e. the characteristic scale imprinted in the early universe remains unaltered. However, because gravitational instability is a nonlinear process, the same motions that lead to the growth of correlations also change their shape. Thus one must check the validity of linear perturbation theory arguments, even at scales as large as $100 \Mpc$. This is even more so given the precision attainable in forthcoming experiments, e.g. a shift in the acoustic scale of one percent generates systematics in the deduced dark energy equation of state parameter $w$ of about five percent~\cite{BAOdetect,Durham2007}, which is not negligible compared to the expected statistical errors in the next generation of galaxy surveys.

Previous work on the subject has been based on numerical simulations~\cite{Durham2007,Nbody,Huff07,ESSS06,Ma2007}, using the halo model~\cite{SW06}, or combinations of analytic and numerical techniques~\cite{komatsu,GuBeSm2006,ESW06,RScube06,MaPie07,RScube07}. The picture emerging from this body of work regarding systematic effects due to nonlinear clustering is still far from converging into a coherent framework.  For example, \cite{SW06} show from the halo model that any clustering-induced systematics must be small, and \cite{ESW06} argue that any systematic shift (i.e. not related to random motions) of the acoustic scale for dark matter in real space must be of unobservably small, of order $0.01\%$ at $z=0$. On the other hand, \cite{GuBeSm2006} showed that the nonlinear power spectrum fitting formula of \cite{Halofit} predicts percent-level shifts in the acoustic peak of the two-point function. Similarly, \cite{RScube06} argue from studying the power spectrum of halos in simulations and perturbation theory that significant shifts are expected. In addition, \cite{Durham2007} do a detailed study of the impact of nonlinearities on the power spectrum of galaxies from semianalytic models to conclude percent-level shifts cannot be excluded. Finally, \cite{RScube07} presents evidence from numerical simulations that the acoustic peak of dark matter and their halos do experience non negligible systematic shifts, which cannot be explained by random motions alone,  and construct a physical model based on the coherent infall of pairs to understand their origin. 

This state of affairs is perhaps not too surprising given that the effects involved are small, and require great accuracy from analytic and numerical methods. In this paper we consider this issue by using renormalized perturbation theory (RPT~\cite{paper1,paper2}), a new approach to follow nonlinear clustering that includes in a systematic way all nonlinear effects in the fluid approximation around a given scale~\footnote{For other, related, recent approaches to understanding nonlinear clustering see~\cite{Valageas06,McDonald07,MaPie07}. In particular, the description of BAO in the power spectrum is also addressed in~\cite{MaPie07}.}. Here we concentrate on fundamental questions such as 1) can nonlinear effects generate shifts in indicators of the acoustic scale large enough that may bias determinations of cosmological parameters? 2) if so, what physics is responsible for this? Is it related to large-scale nonlinearities that we can hope to model accurately, or more complicated physics related to virialized dark matter halos? We shall see that the answer to the first question is `yes', and the answer to the second question involves large-scale physics, that we discuss in detail. Our discussion emphasizes the shifts generated by mode coupling, which constitutes a new result (see also~\cite{RScube07}). In~\cite{paper2} we have already discussed in detail the effects of random motions in terms of large-scale physics; we briefly discuss these here as well in more accessible  terms. That large-scale random motions are responsible for the damping of the linear power spectrum has also been recognized in~\cite{B96,ESW06,ESSS06}.

In the present paper we concentrate on predictions from RPT for the power spectrum and the two-point correlation function. 
A detailed account of the technicalities involved in calculating two-point statistics in RPT and their comparison with numerical simulations is left for a separate publication~\cite{paper4}. Here we present the main results regarding BAO for dark matter in real space and discuss how RPT can shed some light on practical parametrizations of these nonlinear effects in a more general situation when redshift distortions and galaxy bias are also present. No familiarity with RPT  is assumed, the main ideas behind RPT and results on two-point statistics are explained in simple terms in the following section, while the analytic expressions for the power spectrum are presented in Appendix~\ref{RPTpower}.

\section{RPT and Two-Point Statistics}

\subsection{Basics of RPT}
\label{RPTbasic}

Standard perturbation theory (PT, see~\cite{PTreview} for a review) is an expansion of the equations of motion around their linear solution, assuming fluctuations are small. Schematically, for the power spectrum this expansion reads

\beq
P(k,z) = \Dp^2(z)\, P_0(k) + P_{\rm 1loop}(k,z) +P_{\rm 2loop}(k,z) + \ldots
\label{PstdPT}
\eeq
where $\Dp(z)$ is the growth factor at redshift $z$, $P_0(k)$ is the initial power spectrum (at high redshift) so that linear evolution reads $P_{\rm lin}(k,z)=[\Dp(z)]^2 P_0(k)$. In Eq.~(\ref{PstdPT}), $P_{\rm 1loop} \sim {\cal O}(P_{\rm lin}\,\Delta_{\rm lin})$, $P_{\rm 2loop} \sim {\cal O}(P_{\rm lin}\,\Delta_{\rm lin}^2)$, and so on, where $\Delta_{\rm lin} \equiv 4\pi k^3 P_{\rm lin}$ measures the amplitude of fluctuations at scale $k$ in linear theory. For scales approaching the nonlinear regime where $\Delta_{\rm lin} \ga 1$, truncation at any finite order in PT is not meaningful, as neglected higher-order contributions are important. 

In Renormalized Perturbation Theory (RPT,~\cite{paper1}), the main idea is to get around this limitation of PT, by making a resummation of an infinite subset of contributions to the PT expansion. As a result of this process of resummation, where terms of different order have been grouped together into physical objects, what remains is a new series expansion which {\em is not}  a perturbative expansion in the amplitude of fluctuations and, most importantly, exhibits a very different behavior: truncation at finite order in RPT does take into account all nonlinearities from the largest scales down to a given scale, the impact of smaller scales described by the neglected terms is highly suppressed. One the main insights that follows from RPT is that if we write the growth factor as,

\beq
\Dp(z) = \frac{\langle \de_{\rm lin}(\k,z)\, \de_0(\k') \rangle}{\langle \de_0(\k)\, \de_0(\k') \rangle},
\label{Dplus} 
\eeq
where $\de$ denotes the density contrast, and  $\de_{\rm lin}(\k,z)= \Dp(z)\, \de_0(\k)$ is linear evolution (with $\Dp\equiv 1$ at the initial condition); then a whole set of nonlinear contributions to Eq.~(\ref{PstdPT}) (or any correlation function) effectively ``renormalize" the growth factor  to the following, fully nonlinear quantity

\beq
\Dp(z) \longrightarrow G(k,z) = \frac{\langle \de(\k,z)\, \de_0(\k') \rangle}{\langle \de_0(\k)\, \de_0(\k') \rangle},
\label{Prop}
\eeq
where  $\de(\k,z)$ is the fully nonlinear density contrast. The function $G(k,z)$ is known as the {\em propagator}, which can be thought as a measure of the memory of initial conditions, since it gives the time ``propagation" of the cross-correlation between initial and final density contrasts, $\langle \de(\k,z)\, \de_0(\k') \rangle = G(k,z)\, \langle \de_0(\k)\, \de_0(\k') \rangle$. Note that this property means that all the terms in Eq.~(\ref{PstdPT}) that are proportional to $P_0$ (including those in the loop contributions) are resummed into $G^2\, P_0$, whereas in the remaining loop terms the time dependence is dictated by the propagator instead of the growth factor, which essentially means to use Eq.~(\ref{Prop}) to replace the linear propagation in between nonlinear interactions that make up the loop contributions~\footnote{For ease of presentation, we ignore here that the propagator is a matrix, what we call $G$ here is the density propagator for growing mode initial conditions described in~\cite{paper2}. Also, we implicitly assume Gaussian initial conditions, from which the identification of the propagator with the cross-correlation follows. The propagator $G$ has two type of contributions: growing mode and decaying mode. In loop diagrams the latter also become important as after nonlinear interactions perturbations do not remain in the growing mode. See also Appendix~\ref{RPTpower}.}.

The asymptotics of the propagator are easy to understand: at large scales, linear perturbation theory becomes a good approximation and thus 
\beq
G(0,z) = \Dp(z) .
\eeq
On the other hand, at small scales where nonlinear effects are dominant the cross-correlation must be driven to zero, as the final density field resembles very little what it was at the beginning. Thus, we expect on physical grounds that 
\beq
G(k,z) \rightarrow 0\ \ \ \ \  {\rm as\ \ } k \rightarrow \infty.
\eeq
It is this last property that is impossible to capture at a fixed order in PT, where $G$ becomes arbitrarily large in this limit (positive or negative, depending on truncation order), thus leading to unphysical results. 
As we showed in~\cite{paper2}, in the high-$k$ limit the dominant behavior of the propagator can be calculated exactly, giving 
\beq
G(k,z)  \simeq \Dp(z)\, \exp \Big[-{1\over 2}k^2 \sigma_v^2\, \Big(\Dp(z)-1\Big)^2 \Big],
\label{Ghighk}
\eeq
for $k\,\sigma_v\Dp \gg 1$, where the characteristic scale of decay determines the breakdown of linear perturbation theory and it is given by
\beq
\sigma_v^2\equiv\frac{1}{3}\int \frac{P_0(q)}{q^2} d^3q,
\label{sigmav}
\eeq
i.e., the effective one-dimensional amplitude of large-scale velocity flows. Due to the shape of the CDM spectrum of fluctuations, this scale is rather large, at $z=0$
\beq 
[\Dp(0)\sigma_v]^{-1}= 0.15 \kvecMpc 
\label{svz0}
\eeq
while e.g. at $z=1,3$ this becomes $0.24,0.46 \kvecMpc$, respectively. When $k\,\sigma_v\Dp \la 1$ the expression for the propagator is much more complicated, but it is still well approximated by a Gaussian with a weakly scale-dependent width. See~\cite{paper2} for a comparison of the RPT propagator against numerical simulations.

These results have two immediate consequences~\cite{paper2,MaPie07}. First, the validity of linear perturbation theory is much more restricted than commonly thought, as we discuss in more detail below. Second, when such a renormalization is performed, the resulting perturbation theory has the property that large scales are effectively ``shielded" from small scales, e.g. Eq.~(\ref{Ghighk}) suggests that the highly nonlinear regime where the fluid approximation breaks down has an exponentially small impact on large scales, since the scale in Eq.~(\ref{svz0}) is so large compared to scales where more complicated physics is likely to enter. This makes RPT a robust method for probing nonlinear effects at relatively large scales such as those involved in BAO. 

\begin{figure*}[ht!]
\begin{center}
\includegraphics[width=0.49\textwidth]{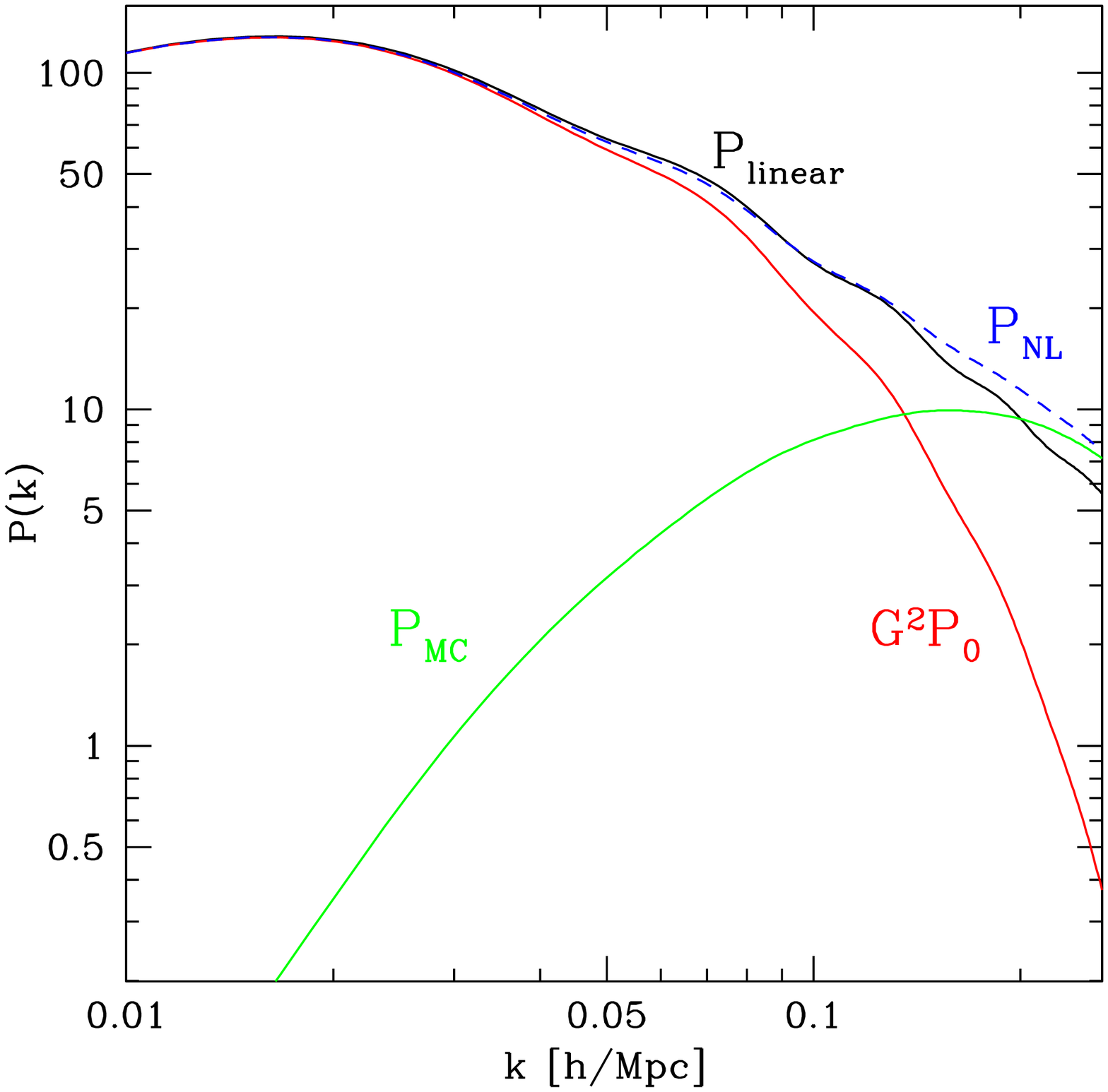}
\includegraphics[width=0.49\textwidth]{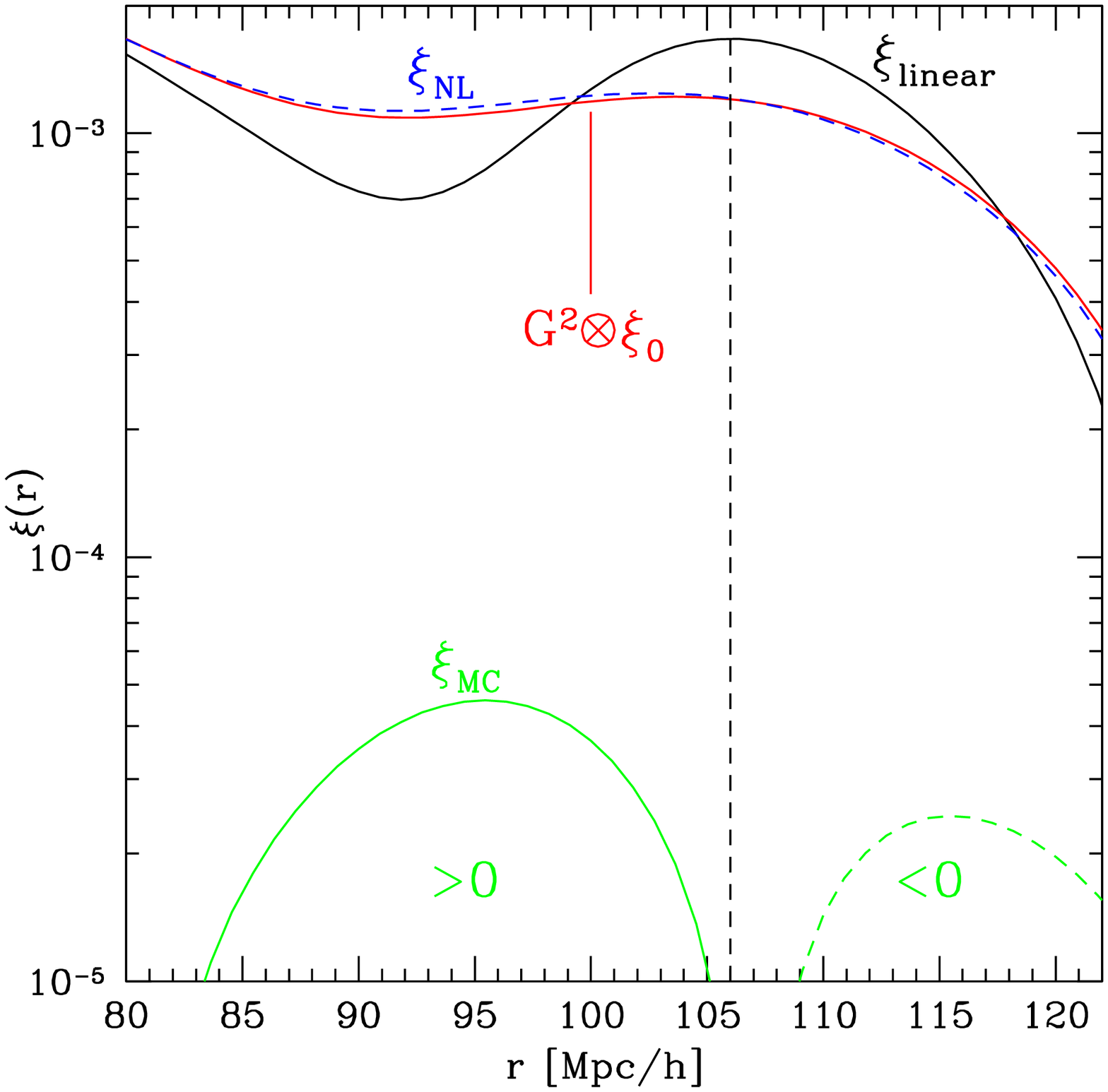} 
\caption{The power spectrum (left) and two-point correlation function (right) as described by RPT, following Eqs.~(\protect\ref{equationP}) and~(\protect\ref{equationXi}), respectively. Each contribution is labeled in the same line styles left and right, and these are just Fourier transforms of each other. The damping of the linear spectrum by $G^2$ (left) translates into a smoothing of the acoustic peak (right), whose unperturbed position is at $106 \Mpc$ in the cosmological model we use in this paper (dashed vertical line). The mode coupling power shows a strong scale dependence (left) and oscillations (of amplitude too small to be seen in this figure, see Fig.~\protect\ref{NodeShift}) and this translates into a shift towards small scales in correlation function space (right). Note that $\xi_{\rm MC}$ is positive left of the peak (solid) and negative right of the peak (dashed). }
\label{pkand2PCF}
\end{center}
\end{figure*}

\subsection{Two-Point Statistics in RPT}
\label{2ptRPT}

When contributions to Eq.~(\ref{PstdPT}) are absorbed into the renormalization of  the growth factor as in Eq.~(\ref{Prop}), the equation for the power spectrum in RPT reduces to an exact but simple form,
\beq
P(k,z)=G^2(k,z) \, P_0(k) + P_{\rm MC}(k,z)
\label{equationP}
\eeq
where {\em all} contributions which are proportional to the initial spectrum of fluctuations are now included in the first term, which  represents how much of the  primordial power remains at a given scale after nonlinear evolution, and thus has direct information on the linear power spectrum. Note that $G$ describes how a given $k$ mode propagates in time in the presence of all the other Fourier modes, and thus it depends on the initial power spectrum at all scales, as it is manifest in Eqs.~(\ref{Ghighk}-\ref{sigmav}). By contrast, the second term $P_{\rm MC}$ represents the power generated by mode-coupling at smaller scales, and depends on the linear power at different scales than $k$ through complicated convolutions. This has a similar loop expansion to that sketched in Eq.~(\ref{PstdPT}), 
\beq
P_{\rm MC}(k,z) = P_{\rm MC}^{\rm 1loop}(k,z) + P_{\rm MC}^{\rm 2loop}(k,z) + \ldots
\label{PmcLoops}
\eeq
but with the important difference that its time dependence is also dictated by the propagator. As a result of this, the convergence properties of the remaining perturbation theory for $P_{\rm MC}$ are drastically changed: only a few terms are needed to describe nonlinear effects at a given scale~\cite{paper1}. This is clear from the physical interpretation of each term in Eq.~(\ref{PmcLoops}). A term with $n$ loops describes the effect of $(n+1)$ modes coupling to the mode of interest, i.e. $\k=\k_1+\ldots+\k_{n+1}$, and thus  $P_{n{\rm loop}} \sim {\cal O}( \Delta^n)$ while in addition there is an exponential suppression at high-$k$ because the time dependence of  $P_{n{\rm loop}}$ is built out of $G$'s. Thus each successive term is increasingly suppressed at  low-{\it k} where $\Delta\ll1$, then peaks at a higher $k$ where $\Delta\ga1$, before decaying when $G\rightarrow 0$. See Section~\ref{nlps} for more discussion on this, and also Fig.~\ref{modecouplingpower} below for how the first two terms in Eq.~(\ref{PmcLoops}) describe the large-scale properties of $P_{\rm MC}$.

The two-point function follows from Eq.~(\ref{equationP}),
\beq
\xi(r,z)= [G^2 \otimes \xi_0](r,z) + \xi_{\rm MC}(r,z),
\label{equationXi}
\eeq
where the symbol $\otimes$ indicates a convolution. Since in Fourier space the propagator $G$ is approximately Gaussian, the first term convolves the initial correlation function with an approximately Gaussian kernel, leading to a smoothing of any features present in $\xi_0$, that also induces shift of the peak location due to the fact that the linear correlation function is not quite symmetric about the peak. The second term, due to mode-coupling, naturally leads to a shift of the acoustic peak in the two-point function, as we will discuss in detail in Section~\ref{APshift}.

Figure~\ref{pkand2PCF} illustrates these results at $z=0$, for the power spectrum  (left panel) and the two-point correlation function (right panel). The acoustic signature in the linear spectrum consists of a roughly harmonic sequence of peaks and troughs exponentially damped due to Silk diffusion. The damped sequence extends up to $k\sim 0.2 \kvecMpc$  while the first trough is around $k\approx0.05\kvecMpc$. In the correlation function the acoustic signature is a single broad peak at the sound horizon scale  (shown also as a vertical dashed line, at $106 \Mpc$ for the cosmology we consider here), and its width is related to the decreasing envelope of oscillations in the power spectrum~\cite{ESW06}. 

Figure~\ref{pkand2PCF} clearly shows the ingredients of RPT at work, following the decomposition in Eqs.~(\ref{equationP}) and~(\ref{equationXi}). The left panel shows the damping of the linear features from $G^2P_0$ as nonlinear motions decorrelate the Fourier modes with respect to their initial values, and the sharp rise of the mode-coupling contribution $P_{\rm MC}$ describing the newly generated power that dominates at small scales. One important conclusion from this is that the validity of linear theory is much more restricted than commonly thought~\cite{paper2,MaPie07}: the comparison between linear and nonlinear power gives a misleading picture. Deviations from linear theory of order 10\% in power happen at scales as large as $k=0.05 \kvecMpc$, and e.g. at $k= 0.13 \kvecMpc$ where linear and nonlinear power differ by less than $2\%$, RPT tells us that approximately {\em only  half} of the power at that scale is related to the primordial power at the same scale, the other half is coming from other scales (see Fig.~\ref{PDet} below for more discussion on this).  Note in addition that this compensation in power between the decay of $G$ and the growth of $P_{\rm MC}$ is only approximate since these terms depend differently on the initial power spectrum and thus cosmological parameters. 

In real space (right panel), the decomposition of nonlinear clustering effects in Eq.~(\ref{equationXi}) is very helpful in providing a qualitative understanding of the cosmological evolution of the acoustic signature. The smoothing of the peak by convolution with $G^2$ can be understood from the Fourier picture by the fact that a peak in real space requires ``well prepared" Fourier coefficients, and as these are decorrelated by non linear evolution the peak decays in amplitude compared to linear amplification. This decorrelation can be understood in real space as resulting from the transport of matter from one point in space to another by large-scale flows: indeed, the result in Eq.~(\ref{Ghighk}) was obtained by resumming exactly the effects of the $\v \cdot \nabla$ terms in the equations of motion~\cite{paper2}. This explains why the characteristic scale of decay in the propagator is the variance of the velocity field, Eq.~(\ref{sigmav}). 

Finally, note that $\xi_{\rm MC}$ changes sign at the acoustic scale, leading to a shift in the peak of the two-point function to smaller scales. This is due to the fact that $P_{\rm MC}$ contains information about the acoustic scale. Indeed, as we show in section~\ref{secS}, $P_{\rm MC}$ contains acoustic oscillations, although of an amplitude too small to be visible in Fig.~\ref{pkand2PCF}.

\section{RPT vs. Numerical Simulations}

\subsection{Numerical Simulations}
\label{nbody}

\begin{figure*}[ht!]
\begin{center}
\includegraphics[width=0.45\textwidth]{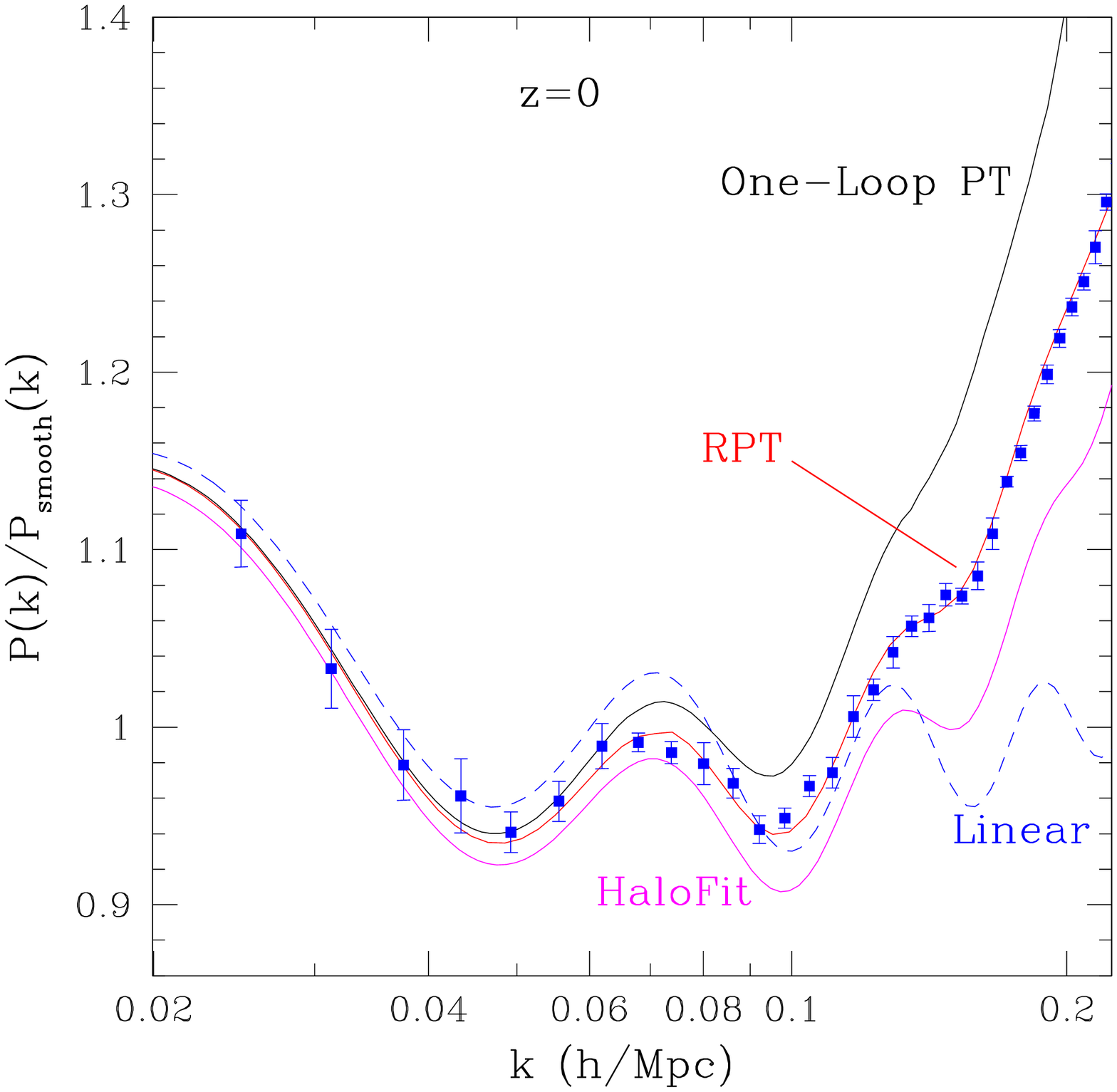}
\includegraphics[width=0.45\textwidth]{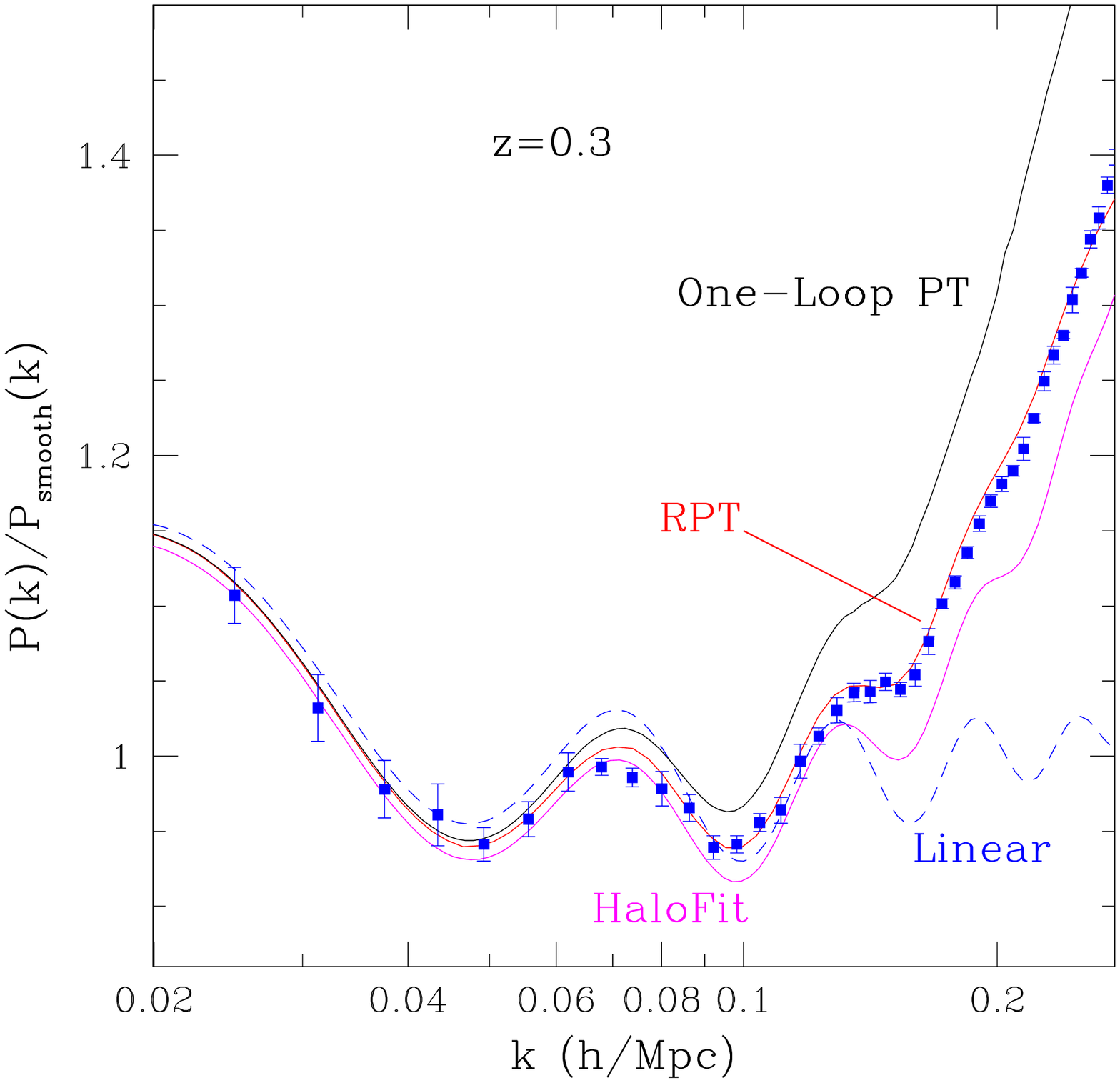} \\
\includegraphics[width=0.45\textwidth]{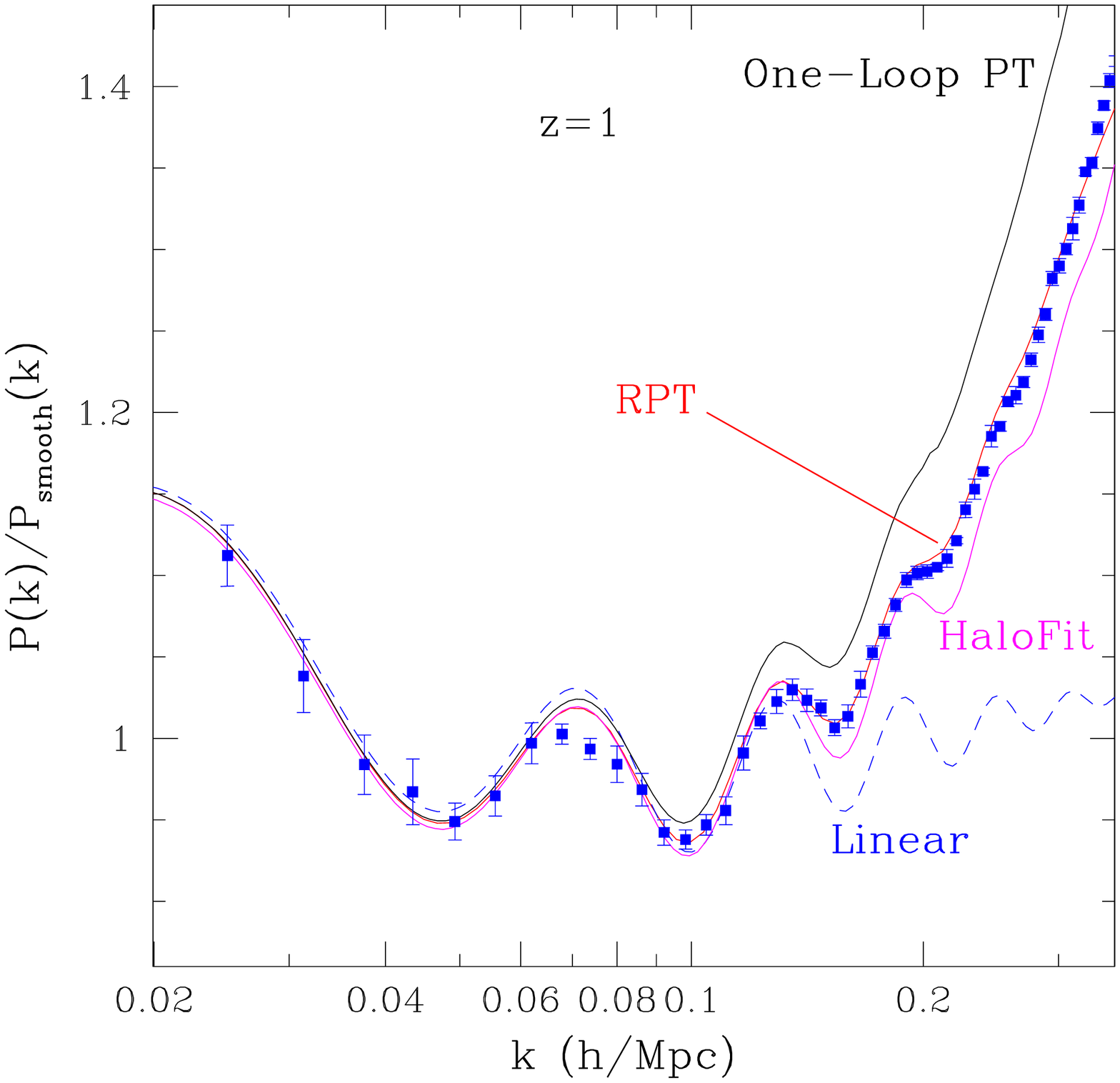}
\includegraphics[width=0.45\textwidth]{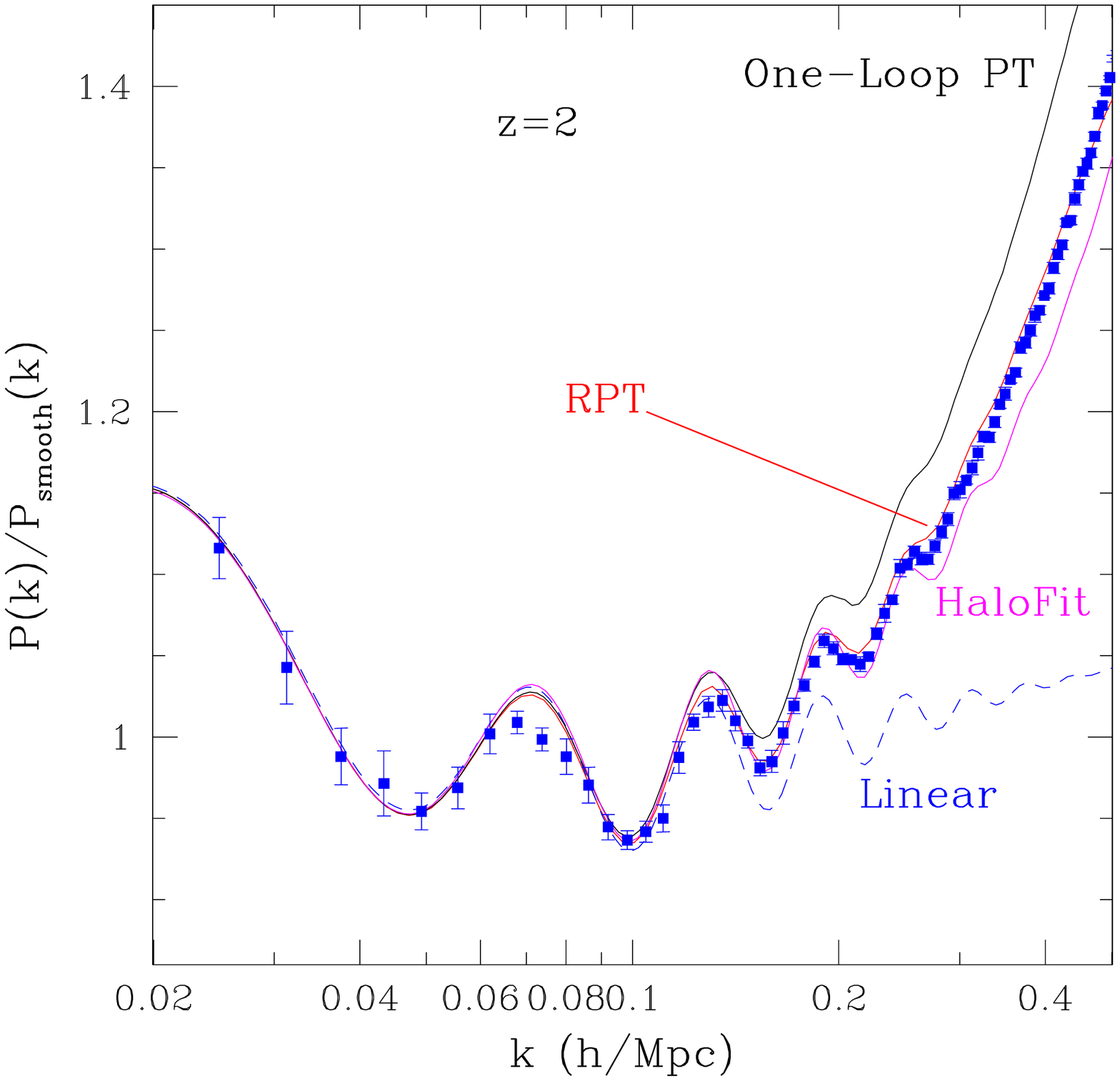}
\caption{Nonlinear evolution of the acoustic oscillations in the dark matter power spectrum. In all cases we show the nonlinear power spectrum divided by a smooth spectrum \cite{bbks} to make the acoustic oscillations more visible. The square symbols with error bars correspond to measurements in N-body simulations, whereas RPT prediction is represented by a solid red line as labeled. One-loop Perturbation theory (solid black line), {\ttfamily halofit} (solid magenta line) and linear theory (dashed blue line) are also shown. The different panels correspond to $z=0,0.3,1,2$ (top left, top right, bottom left and bottom right respectively). The agreement between the RPT prediction and the N-body measurements is excellent for all redshifts, see Fig.~\ref{PDet} for a more detailed comparison.}
\label{nlacoustic}
\end{center}
\end{figure*}

In order to quantify the accuracy at which RPT can describe the evolution of BAO we now proceed to test it against measurements  in N-body simulations. 
A precise description of BAO is somewhat challenging for numerical simulations because this signature lies at large scales of order $\sim 100 \Mpc$. Therefore a very large simulation volume is required to reduce the cosmic variance to levels where it can be dissentagled from true nonlinear evolution, also allowing the oscillations in the power spectrum or the peak in the correlation function to be well resolved. Hence we run a set of $50$ N-body realizations in a cubic volume of side $L_{box}=1280 \Mpc$ and $N_{par}=640^3$ particles with outputs at $z=1,0.5,0$. This gives us a total sampled comoving volume of $\sim 105 (\Gpc)^3$. Initially in this project we used a smaller set of 8 realizations with volume $(1024\Mpc)^3$ and $N_{par}=512^3$ particles with outputs at $z=2,1,0.3,0$, which we only use here in Fig.~\ref{nlacoustic} to cover a larger range in redshift.

The initial conditions were set at $z_i=49$ using the public $2$nd order Lagrangian Perturbation Theory (2LPT) code described in \cite{transients1,transients2}, and the subsequent evolution was followed using the {\sf Gadget2} code \cite{gadget2}. Note that we used the transfer function output at $z=49$ instead of $z=0$, this means our acoustic peak is somewhat smaller in amplitude than it should be. We comment on this issue in section~\ref{APshift}. The cosmological parameters were set to $\Omega_m=0.27$, $\Omega_{\Lambda}=0.73$, $\Omega_b=0.046$ and $h=0.72$. The initial power spectrum had a scalar spectral index $n_s=1$ and was normalized to give $\sigma_8=0.9$ when linearly evolved to $z=0$. We measured the propagator, power spectrum and correlation function at various redshift outputs. In all cases the errors reported in this paper are for  the mean of the ensemble, i.e. corresponding to a volume of $\sim 105 (\Gpc)^3$, and obtained from the scatter among the 50 realizations. Except in Figure~\ref{nlacoustic} where we use the smaller set of simulations, with 8 realizations giving a total volume of $\sim 8 (\Gpc)^3$.

In this paper we focus on a single cosmological model. A more detailed assessment of RPT and numerical simulations, including different cosmologies, will be presented elsewhere~\cite{paper4}. Note however than in~\cite{paper2} we validated the RPT predictions for the propagator for different redshifts and both densities and velocities. Since the evolution of CDM perturbations is not self-similar, this was a non-trivial test. Here we concentrate on the density power spectrum at different redshifts. Higher-order statistics will be considered in a forthcoming paper~\cite{paper5}.

\begin{figure*}[ht!]
\begin{center}
\includegraphics[width=0.49\textwidth]{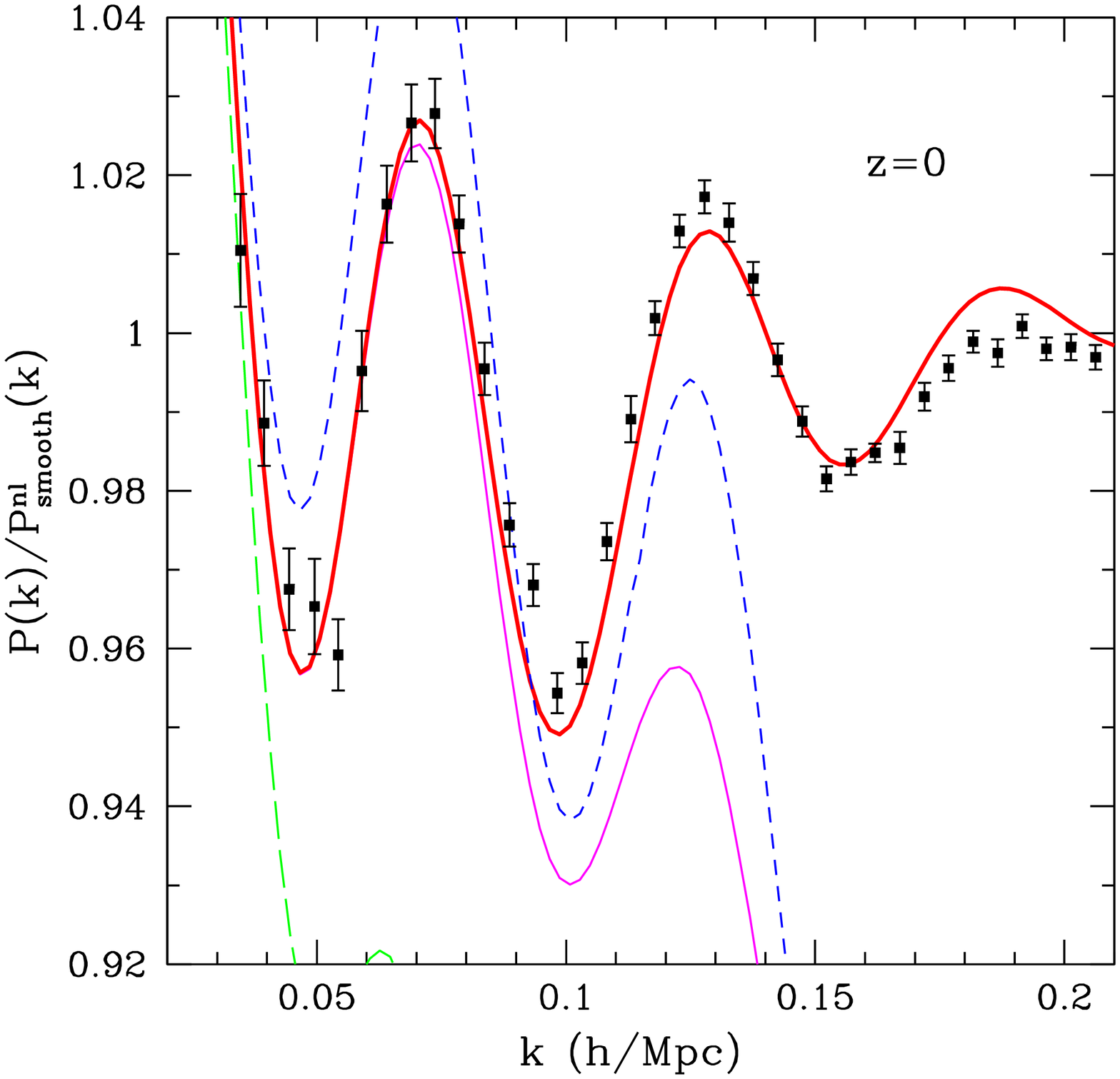}
\includegraphics[width=0.49\textwidth]{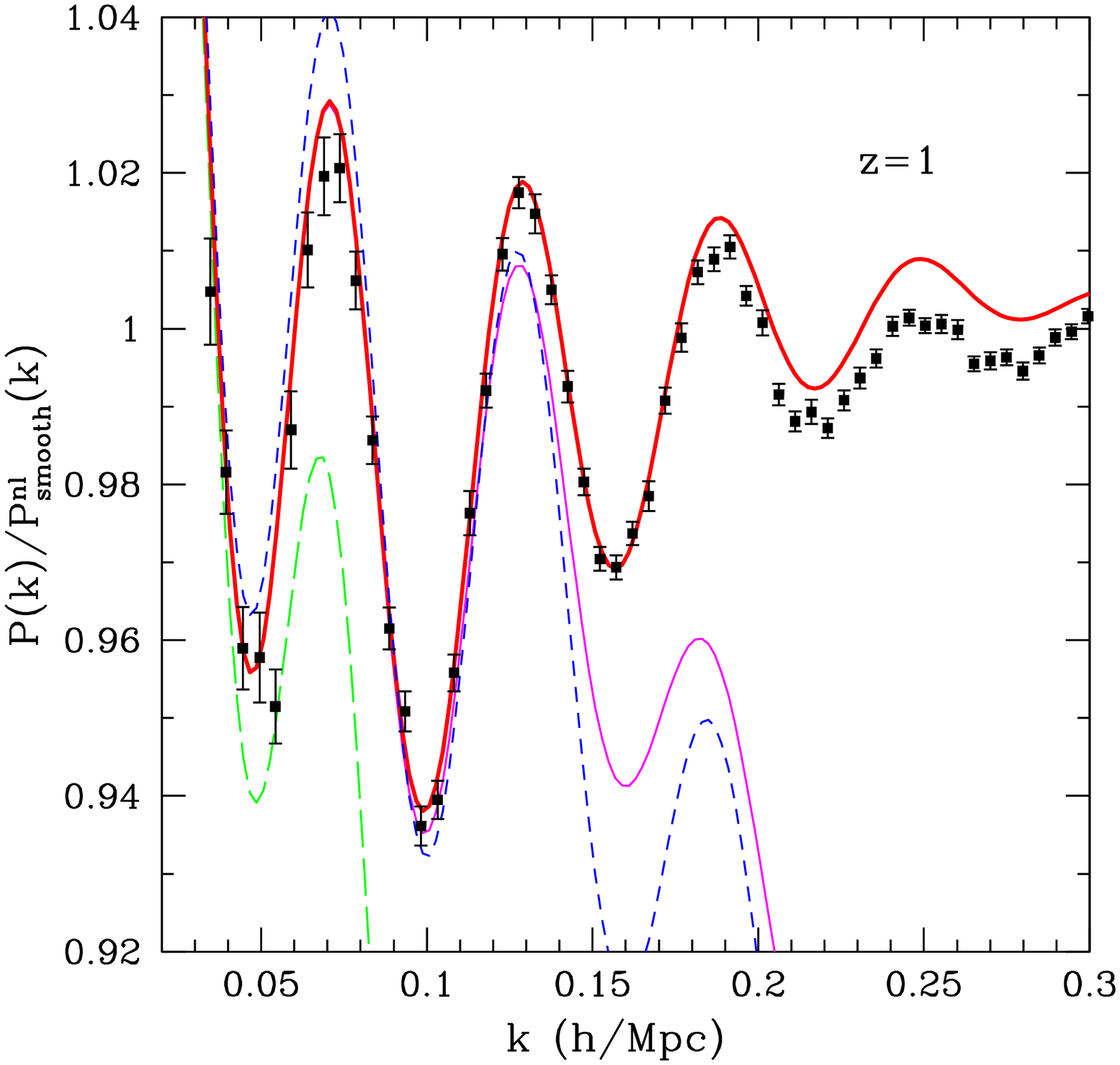} 
\caption{As Fig.\protect\ref{nlacoustic}, but here we divide by the {\em nonlinear} counterpart of $P_{\rm smooth}$ to take out the scale dependence induced by mode coupling and thus provide a more detailed comparison of RPT predictions against simulations. Short-dashed lines denote linear theory, long-dashed lines denote $G^2P_0$, thin solid lines $G^2P_0+P_{\rm MC}^{\rm 1loop}$, and thick solid lines $G^2P_0+P_{\rm MC}^{\rm 1loop}+P_{\rm MC}^{\rm 2loop}$. For $k\protect\ga 0.2 \kvecMpc$ at $z=0$ and $k\protect\ga 0.3 \kvecMpc$ at $z=1$,  the missing higher than two-loop contributions in the mode-coupling power, Eq.~(\ref{PmcLoops}), become important. RPT and simulations agree to better than $1\%$.}
\label{PDet}
\end{center}
\end{figure*}

\subsection{The Power Spectrum}
\label{nlps}

We now present results for the evolution of the baryon wiggles in the power spectrum using RPT and compare them to numerical simulations. According to Eq.~(\ref{equationP}) one needs to calculate the nonlinear propagator $G$ and the mode coupling power $P_{\rm MC}$. Here we explain the basic results in words and refer the reader to Appendix~\ref{RPTpower} for more details about the calculation; a throughout discussion of the technical aspects involved will presented in~\cite{paper4}.

In \cite{paper2} we obtained an analytical prescription for the nonlinear propagator. This was done as follows. First, we  calculated its low-$k$ behavior from one-loop PT and its large-$k$ limit, Eq.~(\ref{Ghighk}), by resumming the infinite subset of diagrams that provided the dominant contribution. We then noticed that these two limits can be matched in a unique way, and without introducing any free parameters, if one regards the low-$k$ limit as an expansion of the Gaussian in Eq.~(\ref{Ghighk}). This procedure resulted in a propagator, for both density and velocity divergence fields, that was shown to be in very good agreement with simulations for all times and scales.

The mode coupling power is decomposed as an infinite sum of  partial contributions, Eq.~(\ref{PmcLoops}), each of which is dominant in a narrow range of scales and negligible otherwise. Roughly speaking, as discussed above, each contribution is a positive ``bump'' centered at increasingly higher $k$ \cite{paper1}, see Fig.~\ref{modecouplingpower} below for the first two contributions. Since the BAO do not span a large range of scales, only few terms in the mode-coupling power are needed to describe them accurately. Indeed, as we shall see in more detail below, the first two contributions describing two and three-mode coupling suffice; i.e. $P_{\rm MC}\simeq P_{\rm MC}^{\rm 1loop}+ P_{\rm MC}^{\rm 2loop}$ is all we need in Eq.~(\ref{PmcLoops}).

Figure~\ref{nlacoustic} shows our results for the dark matter power spectrum as a function of scale for redshifts $z=0,0.3,1,2$  in four panels (as labeled). The symbols with error bars correspond to measurements in 8 realizations of  a cubic box of side $1024 \Mpc$. In addition we also show the predictions from standard one loop PT~\cite{loopPower} and from the nonlinear power spectrum  fitting formula {\ttfamily halofit}~\cite{Halofit}. The dashed line shows the linear power spectrum. Note that all quantities have been divided by a smooth BBKS power spectrum~\cite{bbks} with shape parameter $\Gamma=0.15$ to increase the contrast of the acoustic signature.

We can see that the RPT prediction matches the measurements from the simulations very well at all redshifts. Halofit underestimates the power at all scales and becomes a better fit at higher redshift where nonlinearities are weaker. At $z=0$ the underestimate is about $4\%$ at $k=0.1 \kvecMpc$ and $8\%$ at $k=0.2 \kvecMpc$. As is well known, the prediction from one-loop PT  overestimates the power spectrum, at $z=0$ by about $4\%$ ($18\%$) at $k=0.1 \kvecMpc$ ($0.2 \kvecMpc$). Although this prediction improves at higher redshifts it never becomes accurate, e.g at $z=1$ and $k=0.2 \kvecMpc$ the overestimate is of order $7\%$. This statement might be in conflict with the conclusions in~\cite{komatsu}, where a better than $1\%$ agreement is claimed at these scales and $z\ge1$. The difference may be due to our higher $\sigma_8$ value that makes nonlinearities stronger.

The measurements presented in Fig.~\ref{nlacoustic} show an anomaly at $k \simeq 0.07 \kvecMpc$, where the measured power is significantly below the predictions. We can trace this to the initial conditions of one of the eight realizations. This shows that even in such a large volume $\sim 8 (\Gpc)^3$ the mean power may still differ from the expected cosmic mean at BAO scales.
In what follows we use 50 realizations of $640^3$ particles in a cubical box of side $1280 \Mpc$ that cover a much larger volume $\sim 105 (\Gpc)^3$ and effectively eliminates this issue. Based on the smaller set of simulations~\cite{RScube06} claim that the suppression of the nonlinear power spectrum with respect to linear theory is of order $5\%$ at $z=0$. Based on the much larger volume simulations, we see a maximum suppression of $3.5\%$ at $k=0.07 \kvecMpc$ for the same cosmology.

Figure~\ref{PDet} shows a different normalization than Fig.~\ref{nlacoustic}, this time dividing by the nonlinearly evolved smooth power spectrum, thus taking out the scale dependence induced by the mode-coupling power. This allows us to reduce the vertical scale of the plots and appreciate in more detail the comparison between RPT and simulations. We show linear theory (short-dashed line) and the different contributions that enter into our RPT calculation: $G^2P_0$ (long-dashed line), $G^2P_0+P_{\rm MC}^{\rm 1loop}$ (thin solid line), and  our most complete calculation $P\simeq G^2P_0+P_{\rm MC}^{\rm 1loop}+P_{\rm MC}^{\rm 2loop}$ (thick solid line). For $k\protect\ga 0.2 \kvecMpc$ at $z=0$ and $k\protect\ga 0.3 \kvecMpc$ at $z=1$,  the missing higher than two-loop contributions in the mode-coupling power, Eq.~(\ref{PmcLoops}), become important. The nonlinear smooth power was computed at the same level of approximation, $P= G^2P_0^{\rm smooth}+P_{\rm MC}^{\rm 1loop}+P_{\rm MC}^{\rm 2loop}$, where for the propagator and mode-coupling power we used those of the initial spectrum with wiggles. This minimizes the impact of small numerical integration errors that may affect slightly differently  the smooth and baryonic spectrum. In addition, these ingredients depend weakly on detailed features as they involve integrations over broad ranges of wavenumbers. In detail, though, nonlinear corrections are sensitive to features in the linear spectrum, see discussion in section~\ref{secS} where we show that the mode-coupling power contains oscillations.   We postpone until then a discussion of the shifts in the acoustic pattern of the power spectrum.

From this comparison we see that our calculations agree with simulations to better than $1\%$. 
At such level of detail, however, some of our approximations within the framework of  RPT are likely to matter, see Appendix~\ref{RPTpower}. Similarly, as far as we know simulations have not been tested to this level  so far (see e.g.~\cite{LA85}). Note that testing at this level at these scales requires large simulation boxes, for example according to RPT at $k=0.2 \kvecMpc$ a simulation in a box of size $500 \Mpc$ will have a systematic error of $1\%$ due to its finite volume. The level of agreement in Fig.~\ref{PDet} is nevertheless very encouraging. Note in particular that the thin-solid line ($G^2P_0+P_{\rm MC}^{\rm 1loop}$) is a good approximation to the behavior of the first oscillation, whereas the addition of the two-loop contribution helps to describe the higher harmonics. This is important for the description of the acoustic peak in the two-point function, which is mostly sensitive to the fundamental oscillation. 

\begin{figure*}[ht!]
\begin{center}
\includegraphics[width=0.497\textwidth]{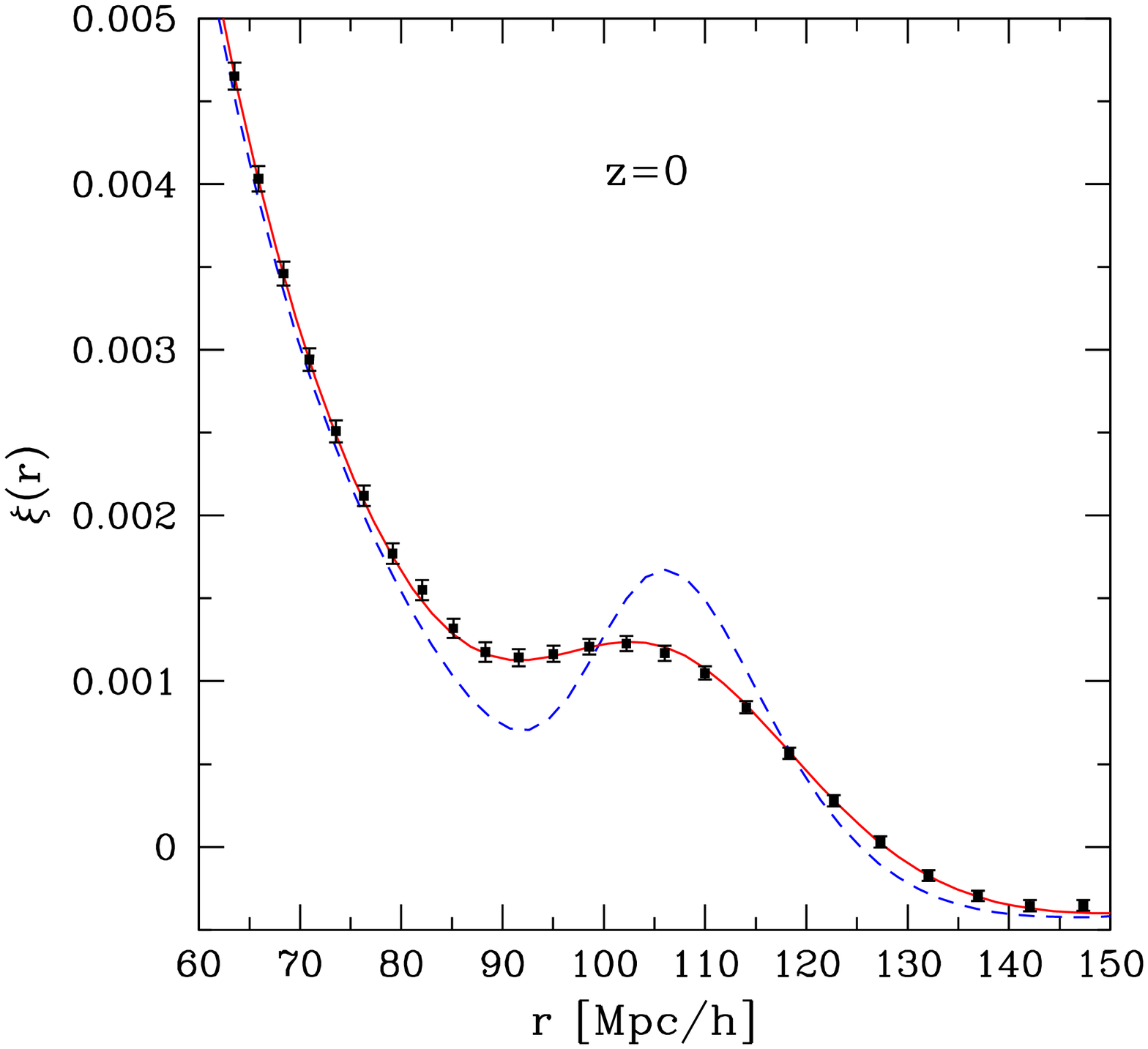} 
\includegraphics[width=0.497\textwidth]{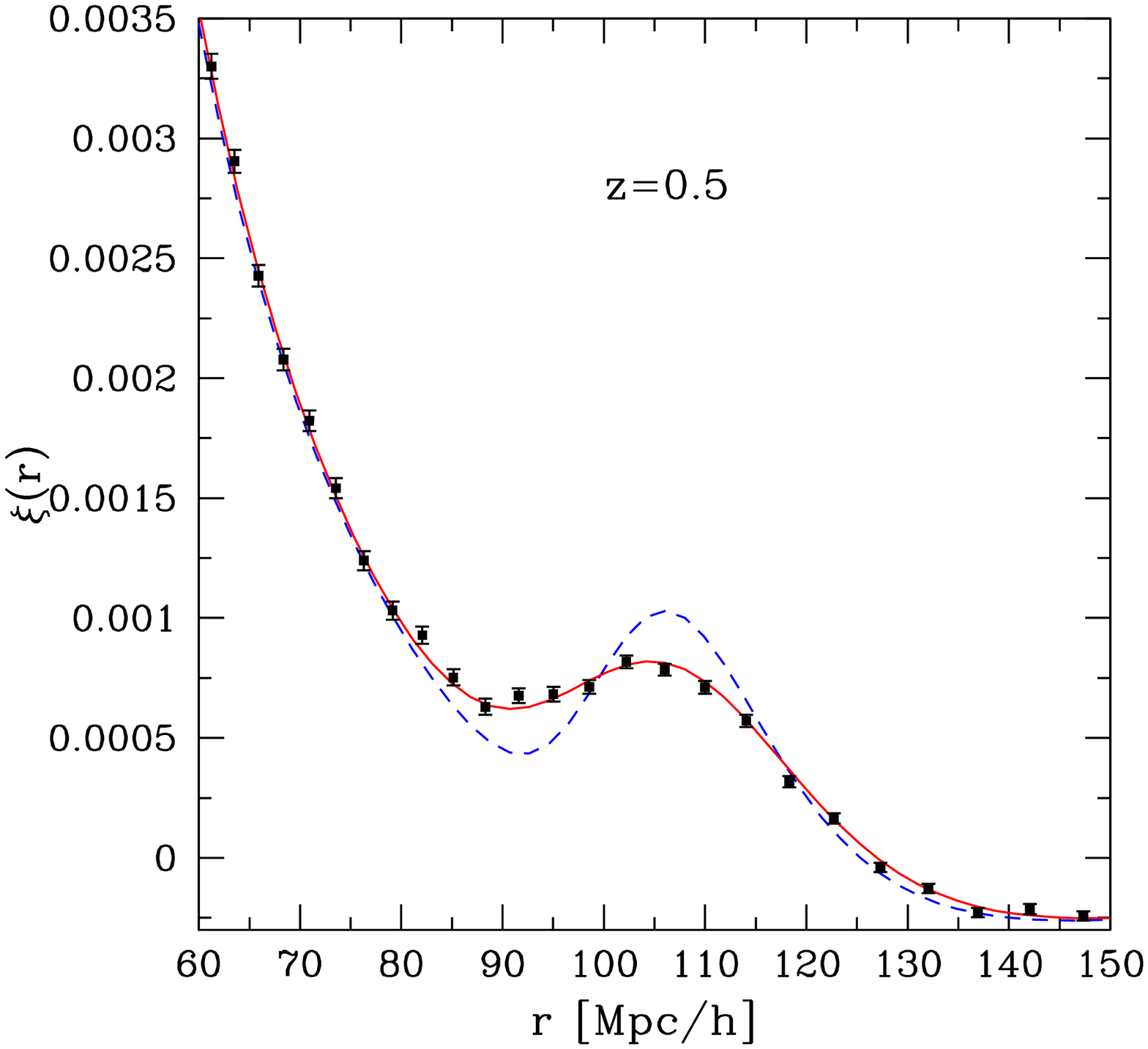}
\caption{The two-point correlation function for $z=0$ (left panel) and $z=0.5$ (right panel). The dashed line corresponds to the linear two-point function, the solid line is the prediction of RPT and the symbols with error bars are the measurements in numerical simulations, corresponding to 50 realizations comprising a total volume of $105~(\Gpc)^3$ approximately.}
\label{xinl}
\end{center}
\end{figure*}

From Figs.~\ref{nlacoustic} and~\ref{PDet} the two effects described in Eq.~(\ref{equationP}) are clearly visible. At lower redshifts more nonlinear evolution increases the damping of the acoustic signature and also increases the scale-dependence induced by  mode-coupling. These deviations from linear theory are substantial, as discussed in Section~\ref{2ptRPT} above. For example, at $z=0$ the correction to linear evolution at $k=0.05 \kvecMpc$ is already about $10\%$ in power, and while at  $k=0.13\kvecMpc$ linear and nonlinear power agree to better than $2\%$, only about half of the nonlinear power is coming from power present in the initial spectrum at the same scale, i.e. $G^2P_0 \approx P/2$.

\subsection{The Two-Point Correlation Function}
\label{nltwopt}

To calculate the two-point correlation function in RPT we Fourier transform the RPT power spectrum prediction presented in the previous subsection, although here we only use the one-loop approximation to the mode-coupling power, i.e. we transform $P \simeq G^2\, P_0 + P_{\rm MC}^{\rm 1loop}$ to real space. The two-loop contribution to $P_{\rm MC}$ is not included since it only introduces very small corrections at BAO scales, and performing its Fourier transform requires a very accurate evaluation, which is numerically costly.

Figure~\ref{xinl} shows the prediction of RPT for the two-point correlation function (solid line) against the measurements in N-body simulations (symbols with error bars) and the linear theory correlation function (dashed line) for a broad range of scales.   The left panel shows $z=0$ and the right panel corresponds to $z=0.5$.

The agreement between RPT and N-body measurements for the two-point function is remarkable, although expected from the results on the power spectrum presented in Figs.~\ref{nlacoustic} and~\ref{PDet}. The different actions of $G^2$ and $\xi_{\rm MC}$ in real space is another way of seeing that the cancellation in the power spectrum between $G^2P_0$ and $P_{\rm MC}$ presents a somewhat misleading picture. The action of these two effects is completely different in the correlation function, and one clearly sees large ($\simeq 30\%$) deviations from linear theory at $100 \Mpc$ scales. This is because measuring the power spectrum at a given scale doesn't say how much of it is correlated with the initial conditions and how much is due to mode-coupling, unless one also measures $G$, something one cannot do in observations but it is simple enough to do in simulations (see~\cite{paper2,paper4} and Section~\ref{RPTvsH} below). As we demonstrate in section~\ref{WS}, the mode-coupling power leads in correlation function space to contributions which involve derivatives of the linear correlation function, and when features are present these terms can become important at large scales. Similarly, convolution with $G^2$ becomes a significant nonlinear effect when the linear correlation function has features which have a width comparable or smaller than that of $G^2$ (given by $2\sigma_v$), which from Eq.~(\ref{svz0}) is $\simeq 13,8,4 \Mpc$ at $z=0,1,3$. The acoustic signature satisfies this condition at low redshift. 

We postpone discussion of the evolution of the peak of the two-point function until section~\ref{APshift}, where we study the shift of the acoustic peak as a function of redshift.

\section{RPT and The Halo Model}
\label{RPTvsH}

\begin{figure*}[ht!]
\begin{center}
\includegraphics[width=0.7\textwidth]{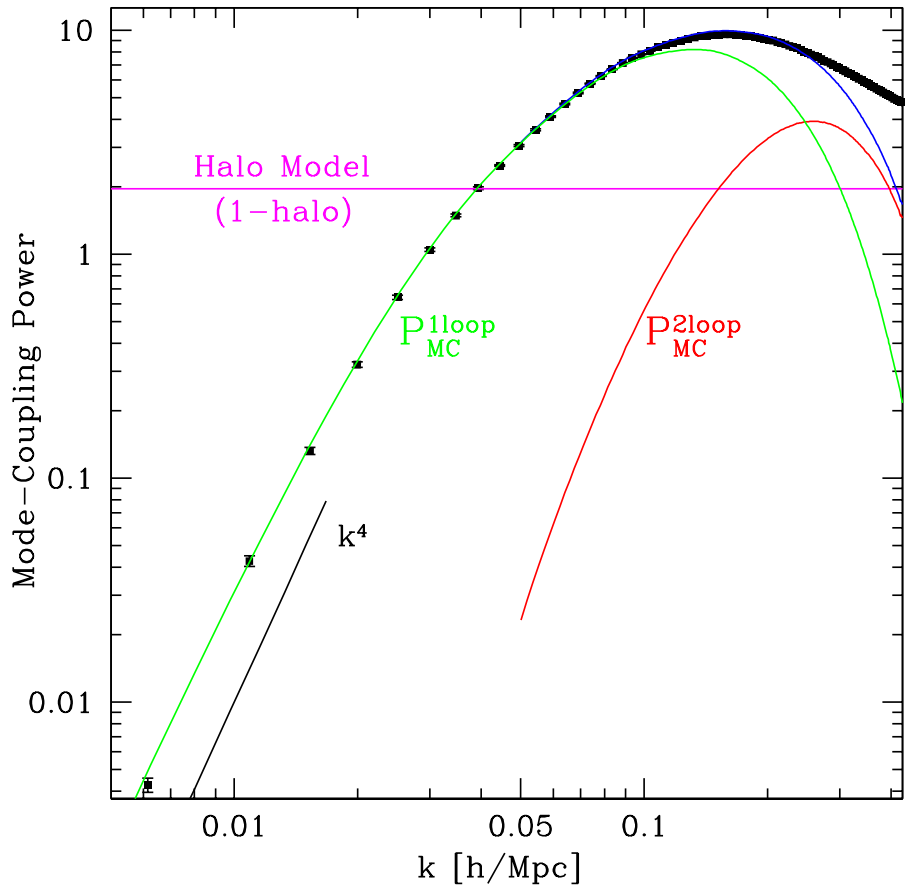}
\caption{The mode-coupling power spectrum $P_{\rm MC}$, see Eq.~(\protect\ref{PmcNb}), measured from N-body simulations (symbols with error bars) as a function of scale at $z=0$. The horizontal line gives the prediction of the halo model, the 1-halo term asymptotics in Eq.~(\ref{HMlowk}), valid at the scales shown in this plot. The other lines give the prediction of $P_{\rm MC}$ in RPT in one and two-loop approximation (as labeled) and their sum (top solid line). For $k \protect\ga 0.2 \kvecMpc$ higher than two-loop contributions not included here become important.  
Note that the low-$k$ asymptote $k^4$ is achieved as predicted by momentum conservation. The halo model grossly violates this behavior.
Although not apparent in this figure, the mode-coupling power contains oscillations, see Fig.~\protect\ref{NodeShift}.}
\label{modecouplingpower}
\end{center}
\end{figure*}

Figure~\ref{pkand2PCF} and Eqs.~(\ref{equationP}) and~(\ref{equationXi}) shows that RPT presents a very similar decomposition to that in the halo model (see~\cite{haloreview} for a review). In this case the density field is modeled as a collection of halos containing all the mass
\beq
\rho(\x)=\sum_i m_i\, u_{m_i}(\x-\x_i),
\label{denhm}
\eeq
where $u_m$ denote the profiles of halos of mass $m$, i.e. how the mass inside each halo is spatially distributed. The density power spectrum can written as
\beq
P(k)=P_{\rm 1h}(k)+P_{\rm 2h}(k) 
\label{halopower}
\eeq
where the ``one-halo" term represents the contribution of objects that are in the same halo (thus dominating at small scales), and the ``two-halo" term dominates at large scales and represents the contribution of pairs in different halos. They are  given by, respectively
\beqa
P_{\rm 1h}(k)\!&\!\!=\!\!& (2\pi)^3\int dm \left( \frac{m}{\bar\rho}\right)^2 n(m)\,  |u_m(k)|^2, \label{onehalo} \\
P_{\rm 2h}(k)\!&\!\!=\!\!& [b_1(k)]^2 \, P_{\rm lin}(k) , \qquad \qquad
\label{twohalo}
\eeqa
with a scale-dependent bias factor
\beq
b_1(k)= (2\pi)^3  \int dm \left(\frac{m}{\bar\rho}\right) n(m)\, b_1(m)\, u_{m}(k), 
\label{bhalo}
\eeq
where $n(m)$ is the mass function, $b_1(m)$ is the linear bias factor for halos of mass $m$ and $u_m(k)$ is the Fourier transform of the halo profile, normalized given Eq.~(\ref{denhm}) so that
\beq
 (2\pi)^3\ u_m(k=0)\ = \int d^3x\ u_m(\x) = 1.
\label{unorm}
\eeq

In~\cite{paper1} we showed that interesting similarities can be established between the halo model and RPT, in particular, the scale dependent bias factor in Eq.~(\ref{twohalo}) was shown to be $b_1(k)=G_h(k,z)/\Dp(z)$ where $G_h$ is the propagator calculated in the halo model following the rules of RPT. 
Written in this simplified and often used form the two-halo term would correspond to propagator renormalization in RPT, i.e. the term $G^2\, P_0$ in Eq.~(\ref{equationP}).  The one-halo term can then be identified with the mode coupling power since it describes mode-coupling in the same sense as $P_{\rm MC}$ in RPT: these terms represent the contribution to the power that is not proportional to the linear spectrum at the same scale.

Here we would like to point out some important differences between RPT and the halo model as written above, which point to serious shortcomings of the latter for BAO modeling. First, in the halo model as written above the characteristic scale of decay of BAO is related to the virial radius of halos through the dependence on the halo profile in Eq.~(\ref{bhalo}), and the resulting spatial scale is {\em much smaller} than that from Eq.~(\ref{svz0}) thus ``delaying'' the decay until higher $k$'s. Furthermore, the decay as a function of scale is only power-law, instead of Gaussian. All this leads to a halo-model propagator or $b_1(k)$ that decays too slowly as $k$ increases. This is not too surprising, as RPT tells us that large-scale nonlinear effects unrelated to smaller scale virialization physics are responsible for the decay of the propagator. The obvious way to improve this situation would be to add nonlinear contributions from both dynamics~\cite{Yang03,Zheng04,Tinker05} and bias~\cite{McDonald06,RScube06} to the two-halo term in order to try to model BAO scales.

The second, related, and more difficult to fix problem is that the one-halo term has a very different behavior at large scales than the mode-coupling power in RPT. In fact, in the low-$k$ limit appropriate for BAO scales using Eq.~(\ref{unorm}) in  Eq.~(\ref{onehalo}) it follows that
\beq
P_{\rm 1h}(0)=\int {dm\over(2\pi)^3} \left( \frac{m}{\bar\rho}\right)^2 n(m) = 1.96\, (\Mpc)^3,
\label{HMlowk}
\eeq
where we have used the Sheth-Tormen mass function~\cite{ST99} at $z=0$. This means that on scales relevant to BAO the halo model power spectrum, Eq.~(\ref{halopower}), reads,
\beq
P(k)\simeq [b_1(k)]^2 P_{\rm lin}(k)+ c,
\label{HMpowLk}
\eeq
where $c=P_{\rm 1h}(0)$. This leads to a very different picture of the impact of nonlinearities on the acoustic signature. Indeed, Eq.~(\ref{HMpowLk}) suggests that while in power spectrum space the baryonic wiggles will be affected by adding a constant power (apart from the too small damping coming from $b_1(k)$ already discussed), the acoustic peak in the two-point correlation function will not experience any shift because the one-halo term is so slowly varying in Fourier space that its Fourier transform will have no structure on $100\Mpc$ scales. In fact, an explicit calculation of the acoustic peak shift in the halo model was not able to find any shift, placing an upper limit of $0.1\%$~\cite{GuBeSm2006}. 

This argument has been made repeatedly in the literature as an explanation of why the acoustic signature should be stable under nonlinearities, see e.g.~\cite{SW06,ESW06}. However, we believe that such arguments are, fundamentally, incorrect. The relevant nonlinear effects are not related to virialized objects, but rather to deviations from linear perturbation theory at much larger scales (see Fig.~\ref{pkand2PCF}) by the same physics that generates a non-zero bispectrum at the largest scales (see section~\ref{WS} below).

Furthermore, we now show that the low-$k$ behavior of the one-halo term is in obvious conflict with simulations.  Figure~\ref{modecouplingpower} shows the measured mode-coupling power $P_{\rm MC}$ in simulations (symbols with errorbars) as a function of scale, for all the large-scale modes available in our simulations (corresponding to a fundamental mode $k=2\pi/1280 \kvecMpc\approx 0.05 \kvecMpc$). These measurements were done by measuring the power spectrum and the propagator, or cross-correlation between initial and final density fields. From Eqs.~(\ref{equationP}) and~(\ref{Prop}) the mode-coupling power can be measured in simulations by calculating $P-G^2P_0$, that is
\beq
P_{\rm MC}(k)\ \delta_D(\k+\k') = \langle \delta(\k) \delta(\k') \rangle - \frac{\langle \delta(\k) \delta_0(\k') \rangle^2}{\langle \delta_0(\k) \delta_0(\k') \rangle},
\label{PmcNb}
\eeq 
where $\delta$ is the final density field, and $\delta_0$ the initial one.

Note that the mode-coupling power at the largest scales is about four orders of magnitude smaller than the linear power (e.g. compare to Fig.~\ref{pkand2PCF}); we are able to measure such small power precisely because of the large volume that our simulations probe. The agreement between the RPT prediction and simulations is impressive, and the failure of the halo model prediction is evident. Clearly, using the low-$k$ limit of the one-halo term to make any predictions is, at best, risky. 

We stress that the RPT prediction has no free parameters whatsoever, see Eq.~(\ref{PmcLowk}) below for an expression valid in the low-k limit. As $k\rightarrow 0$, the prediction is 

\beq
P_{\rm MC}(k,z) \longrightarrow {9\over 98}\  k^4 \, \int \Big[ \frac{P_{\rm lin}(q,z)}{ q^2} \Big]^2 d^3q.
\label{k4tail}
\eeq 
In fact, this low-$k$ asymptote is a well-known result expected from local conservation of momentum: collapsing structures at small scales can only affect large scales through a $k^4$ tail~\cite{Zel65,Peebles80,GGRW86,PTreview}. 

The halo model, on the other hand, predicts that local collapse of dark matter particles into halos of size of order a Mpc can change the dark matter power spectrum significantly on Gpc scales!  It is not clear to us at present how to fix this problem with the one-halo term; although one can certainly generate a $k^4$ tail by properly including mode-coupling physics in the two-halo term, one would, in addition, have to somehow cancel the spurious contribution from the one-halo term. See Section 4.4 in~\cite{haloreview} for more discussion on this.
 
One may wonder whether this discussion is at all relevant for galaxies, since tracers such as galaxies do not conserve momentum because they have significant gravitational interactions with the dark matter and other tracers, the $k^4$ tail argument does not apply. Thus, a contribution to the mode-coupling power from virialized halos, roughly constant on BAO scales, may be present. Although given that the low-$k$ one-halo term prediction is in such serious conflict with simulations for the dark matter, perhaps one should take specific predictions from it for galaxies with a good dose of skepticism. 

Finally we note that the RPT prediction for the mode-coupling power is accurate well beyond the low-$k$ limit. Figure~\ref{modecouplingpower} shows the leading contribution to the mode coupling power in RPT, $P_{\rm MC}^{\rm 1loop}$ (corresponding to the coupling of two Fourier modes), and the next order contribution $P_{\rm MC}^{\rm 2loop}$ (corresponding to the coupling of three Fourier modes), which peaks at smaller scales.  The sum of these two contributions (top solid line) matches the simulation results very well up to $k \simeq 0.2 \kvecMpc$ at $z=0$. For $k \ga 0.2 \kvecMpc$ three-loop (four-mode coupling) and higher order contributions not included here become important.

\begin{figure*}[ht!]
\begin{center}
\includegraphics[width=0.497\textwidth]{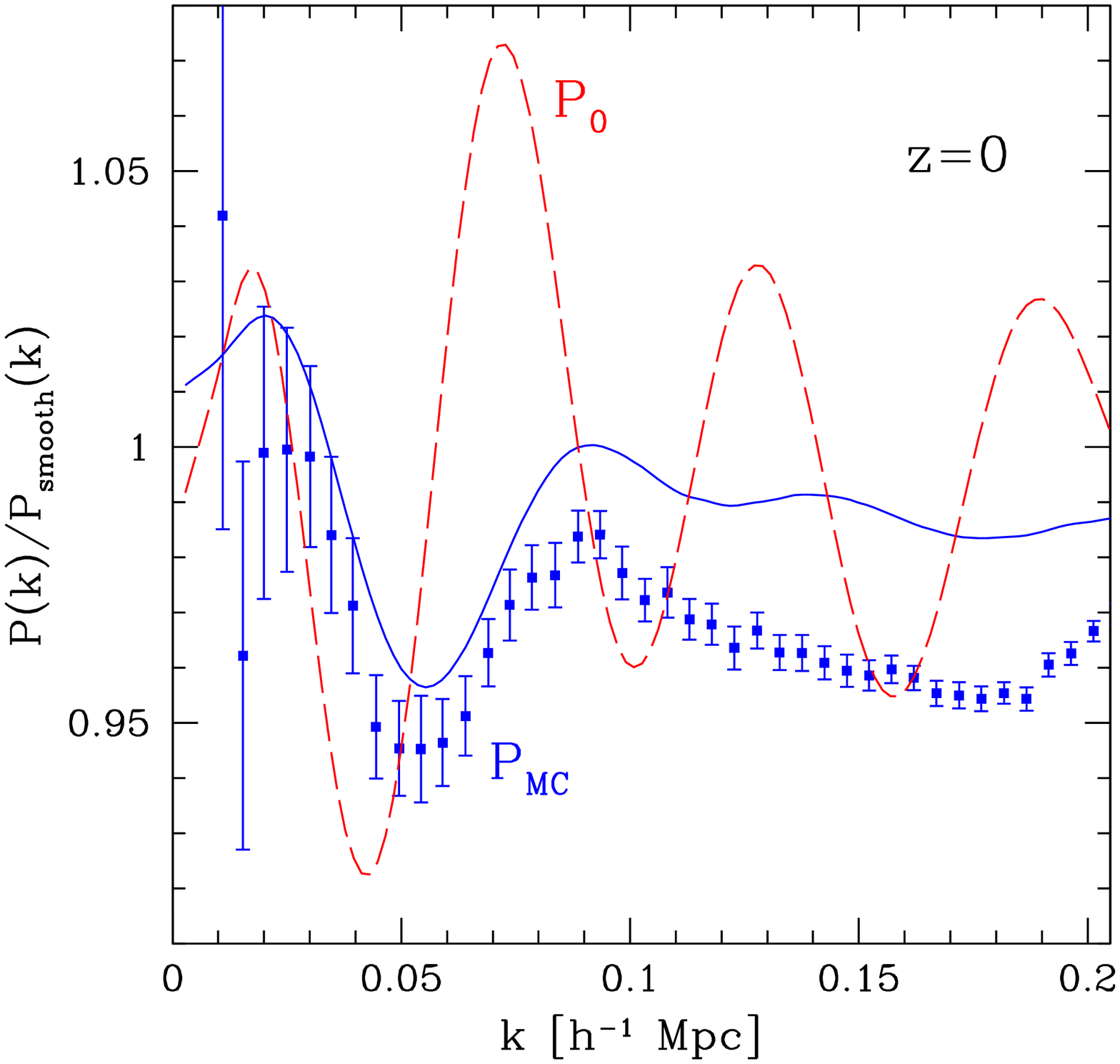} 
\includegraphics[width=0.497\textwidth]{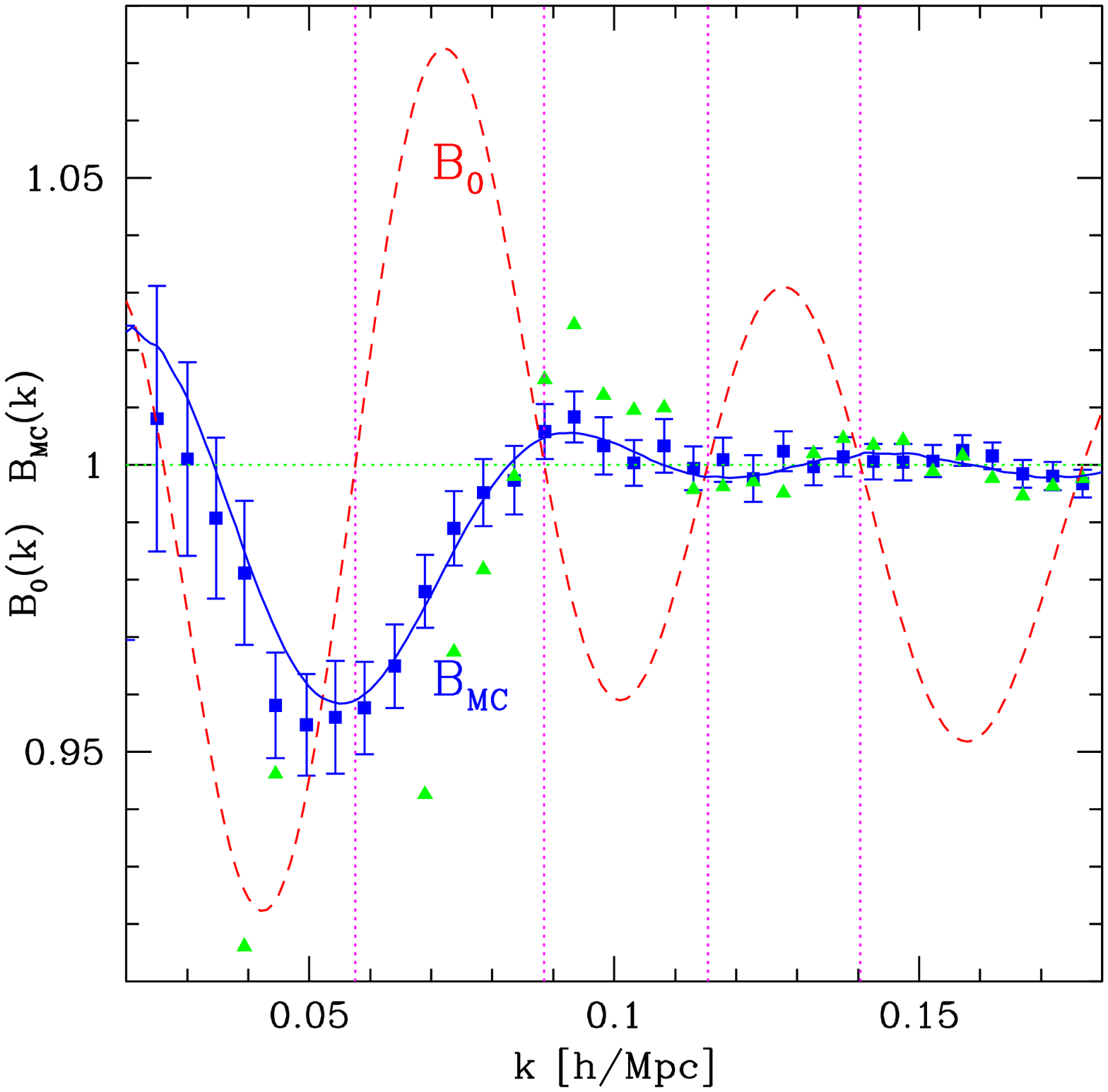}
\caption{{\em Left Panel}. Acoustic oscillations are induced in the mode-coupling power spectrum. The solid line denotes the ratio of $P_{\rm MC}$ to $P_{\rm MC}^{\rm smooth}$, with the latter constructed by evolving a smooth spectrum using RPT. The dashed line corresponds to the same ratio but in linear theory ($P_0$). The presence of acoustic oscillations in $P_{\rm MC}$ is also evident in the measurements in N-body simulations. {\em Right Panel}. The ratios of wiggly to smooth spectra, $B_0=P_0/P_0^{\rm s}$ for linear theory (dashed lines) and $B_{\rm MC}=P_{\rm MC}/P_{\rm MC}^{\rm s}$ for the mode-coupling power in RPT (solid lines) and N-body simulations (symbols with error bars). The triangle symbols (with error bars suppressed for clarity) denote $B_{\rm MC}$ obtained in N-body simulations by the rebinning method. The vertical lines denote four linear spectrum oscillation nodes where $B_0=1$. Note that the oscillations in $B_{\rm MC}$ are {\em out of phase} with respect to $B_0$, leading to shifts in the power spectrum nodes toward higher wavenumbers, see Eq.~(\protect\ref{NLnodes}).}
\label{NodeShift}
\end{center}
\end{figure*}

\section{Shifts}
\label{secS}

We now discuss the issue of how nonlinear evolution may induce shifts in the acoustic signatures, which can potentially bias the determination of the sound horizon and lead to systematics in the determination of cosmological parameters such as the equation of state of the dark energy. The discussion is based on shifts of the power spectrum wiggles and the acoustic peak of the two-point function, which one can roughly think of as indicators for the size of the sound horizon. A  proper assessment of the impact of such shifts on the determination of the sound horizon depends on the procedure used to analize the data, this is beyond the scope of this paper. However, our results have something useful to say on what such procedures need to take into account.

\begin{figure}[ht!]
\includegraphics[width=0.495\textwidth]{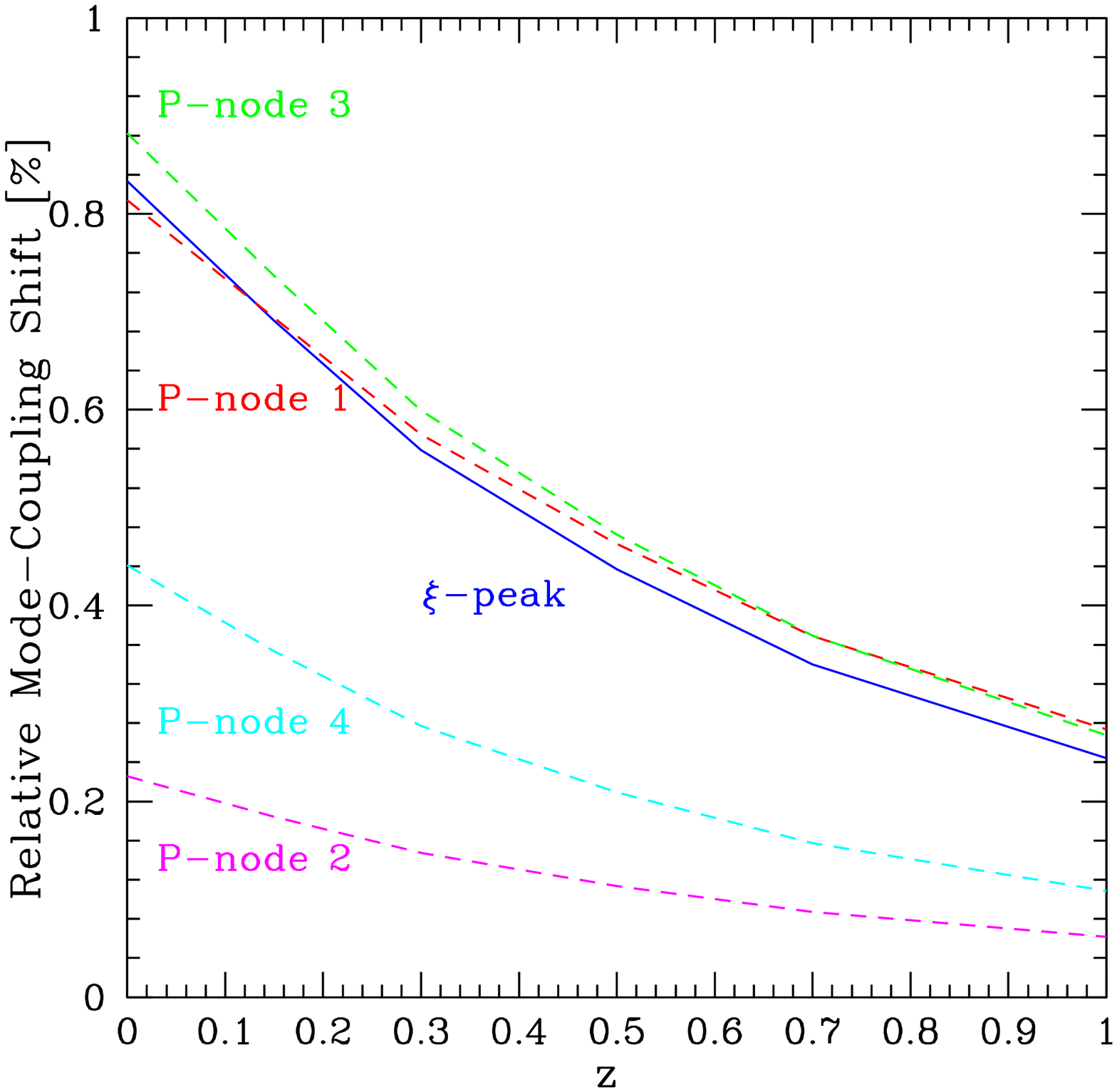}
\caption{Percentual shifts due to mode-coupling as a function of redshift. The solid line denotes the shift in the acoustic peak position of the two-point function. The dashed lines denote the shifts in the power spectrum first four nodes in the method by~\protect\cite{Petal07} (see Fig.~\protect\ref{NodeShift}). Note that this calculation corresponds to dark matter in real space, thus it should be considered as a lower limit when redshift space distortions and galaxy bias are present.}
\label{shiftvsredshift}
\end{figure}

\subsection{Power Spectrum Wiggles}
\label{PSW}

The evolution of the acoustic pattern in the power spectrum, Figs.~\ref{nlacoustic} and \ref{PDet}, is rather complex.
Looking at the position of maxima and minima of the baryon wiggles we can notice some difference between linear and nonlinear cases. While there are definitely shifts between those at the percent level, quoting numbers is dependent on which quantity one is interested in (see e.g.~\cite{Netal07}). 

Here we follow the method described in~\cite{Petal07}, which was recently applied to measuring the BAO scale in the SDSS and 2dfGRS galaxy power spectra. They define a smooth power spectrum out of the observed nonlinear power with acoustic oscillations by doing a spline fit to a coarse rebinning of it. Then they use the ratio of the power to the smooth power and look for oscillation nodes in this ratio that oscillates about unity. These nodes are then used to calculate the sound horizon scale.

We define a smooth nonlinear power spectrum by analogy with Eq.~(\ref{equationP})

\beq
P^{\rm s}(k)=G^2(k)\, P_0^{\rm s}(k)+ P_{\rm MC}^{\rm s}(k),
\label{Psmooth}
\eeq
the idea being that each term is a smooth version of the terms in Eq.~(\ref{equationP}). Note that already the propagator $G$ is a smooth quantity, e.g. due to Eqs.~(\ref{Ghighk}-\ref{sigmav}); thus $P_0^{\rm s}$ is simply a smooth version of the linear spectrum $P_0$. However, we {\em do not} assume that $P_{\rm MC}^{\rm s}$ is equal to $P_{\rm MC}$, i.e. we allow for oscillations in $P_{\rm MC}$; this is crucial in what follows and is why we reach different conclusions than~\cite{Petal07}. 
The ratio of nonlinear powers $B\equiv P/P^{\rm s}$ can then be written as,
\beq
B(k)= g(k)\, B_0(k) + [1-g(k)]\, B_{\rm MC}(k),
\label{Bratio}
\eeq
where $B_0\equiv P_0/P^{\rm s}_0$ is the linear ratio and 
\beq
g(k) \equiv \frac{G^2(k) P_0^{\rm s}(k)}{P^{\rm s}(k)}, \ \ \ \ \ B_{\rm MC}(k) \equiv \frac{P_{\rm MC}(k)}{P^{\rm s}_{\rm MC}(k)}.
\label{gBmc}
\eeq

To measure the acoustic scale~\cite{Petal07} implicitly assume that $B_{\rm MC}(k)=1$, in which case it follows from Eq.~(\ref{Bratio}) that the power spectrum nodes defined as $B(k_{\rm nodes})=1$ coincide with the linear spectrum nodes $B_0(k_{\rm nodes})=1$. Thus, they argue, nonlinear evolution does not shift the values of $k_{\rm nodes}$, which can be related to the acoustic scale, e.g. in the approximation of~\cite{BG03} $k_{\rm nodes}=n\pi/r_s$, where $r_s$ is the comoving sound horizon scale at recombination. 

However, we now show that the assumption $B_{\rm MC}(k)=1$ does not quite hold, leading to shifts in the values of $k_{\rm nodes}$ obtained from linear theory. In order to do so, it remains to define $P_0^{\rm s}$ and $P_{\rm MC}^{\rm s}$ appearing in Eq.~(\ref{Psmooth}). Instead of defining smoothed quantities by rebinning we proceed in a slightly different way, for reasons that will become clear shortly. The idea, to first approximation, is to construct $P^{\rm s}$ out of nonlinear evolution of the linear spectrum $P_0^{\rm s}$.

We define the smooth version of the linear spectrum by modifying the BBKS spectrum that approximates the overall shape of the wiggly power spectrum with analytic functions that are rigorously smooth.  A good check is to ensure that $\int  P_0^{\rm s}(q) dq = \int  P_0(q) dq$, which in turn will lead to a propagator $G$ that is equal for the evolution away from smooth and wiggly initial conditions. The ratio $B_0= P_0/P^{\rm s}_0$ so constructed is shown in the left panel of Fig.~\ref{NodeShift} as dashed lines. We then calculate the mode coupling contribution for the initial spectrum $P_0^{\rm s}$ using RPT, let's call this result $P_{\rm MC}^{\rm smooth}$. This leads to a ratio $P_{\rm MC}/P_{\rm MC}^{\rm smooth}$ that is shown in the left panel of Fig.~\ref{NodeShift} as solid lines. 

As it is obvious from this plot, we see that the mode-coupling power {\em is not smooth}. It contains oscillations that are approximately 90 degrees out of phase with the acoustic oscillations of the linear power (dashed lines). This can be understood from the fact that nonlinear corrections to the power spectrum are a decreasing function of the local spectral index~\cite{SF96}:  the dashed curve in Fig.~\ref{NodeShift} modulates the slowly varying spectral index of the smooth power, one thus expects maximal (minimal) nonlinear corrections at those wavenumbers where the derivative of the dashed line is most negative (positive). This creates an oscillating correction that is out of phase. 

Also shown in the same plot is the ratio of $P_{\rm MC}$ measured in the simulation (symbols with error bars) obtained from the measured power and propagator $P_{\rm MC}=P-G^2\, P_0$ to the smooth mode-coupling power $P_{\rm MC}^{\rm smooth}$ calculated from RPT as discussed above. We see that the same oscillations are seen in the N-body data, although there is a significant discrepancy between RPT and N-body measurements for $k \ga 0.1 \kvecMpc$, more so than in the comparison of total power in Fig.~\ref{PDet}. This is due to the fact that the measured propagator decays as a function of $k$ slightly more slowly than in RPT, thus substracting $G^2\, P_0$ from the measured power leads to a lower $P_{\rm MC}$. 

To finalize our definition of $P_{\rm MC}^{\rm s}$ in Eq.~(\ref{Psmooth}) we note that the ratio $P_{\rm MC}/P_{\rm MC}^{\rm smooth}$ (solid line in the left panel of Fig.~\ref{NodeShift}) does not oscillate about unity as $k$ increases due to a small drift. To fix this and thus have a smooth nonlinear spectrum about which the wiggly spectrum oscillates, it is simply enough to define $P_{\rm MC}^{\rm s}$ as $P_{\rm MC}^{\rm smooth}$ slightly tilted (by $1\%$ at $k=0.2\kvecMpc$) by a linear or mild quadratic dependence on $k$. This brings $B_{\rm MC}$ in Eqs.~(\ref{Bratio}-\ref{gBmc}) to a function that oscillates about unity, as it should. The same procedure can be done for the N-body measurements. The result of this is shown in the right panel of Fig.~\ref{NodeShift}. We have thus constructed smooth spectra $P^{\rm s}$  about which the wiggly spectra oscillate~\footnote{The mode-coupling spectra $P_{\rm MC}^{\rm s}$ are slightly different for RPT than for the N-body simulations, which reflects that the linear spectrum is damped slightly differently in RPT than in the N-body case. That is, the smooth total spectra $P^{\rm s}$ will be very similar (to the level of agreement seen in Fig.~\ref{PDet}) in RPT and in the N-body case as the two terms in Eq.~(\ref{Psmooth}) compensate.}.

We are now ready to see what effect does $B_{\rm MC}$ (solid line in the right panel of Fig.~\ref{NodeShift}) have on the node positions determined from $B(k_{\rm nodes})=1$ in Eq.~(\ref{Bratio}) compared to those in linear theory, $B_0(k^{\rm linear}_{\rm nodes})=1$ (given by the vertical lines in Fig.~\ref{NodeShift}). Because near the linear nodes $B_{\rm MC}$ is smaller (larger) than unity when $B_0$ has positive (negative) derivative, one can easily see graphically that the net result is to shift all nodes to larger wavenumbers. Indeed, close to linear nodes we can write $B_0 \simeq 1+ \alpha\, (k-k^{\rm linear}_{\rm nodes})$ and $B_{\rm MC} \simeq 1-\epsilon$, where the signs of $\alpha$ and $\epsilon$ always coincide. Then Eq.~(\ref{Bratio})  says that the position of the nodes after nonlinear evolution shifts to

\beq
k_{\rm nodes}=k^{\rm linear}_{\rm nodes} + \Big( \frac{\epsilon}{\alpha} \Big) \Big( \frac{P_{\rm MC}^{\rm s}}{G^2P^{\rm s}_0} \Big) 
> k^{\rm linear}_{\rm nodes},
\label{NLnodes}
\eeq
and thus all nodes shift to smaller scales, as expected. Note that as $k$ increases, $\epsilon$ decreases while $\alpha$ slowly decreases; however, the ratio $P_{\rm MC}^{\rm s}/G^2P^{\rm s}_0$ increases rapidly as the mode-coupling power becomes the dominant contribution to the power (see Fig.~\ref{pkand2PCF}). Thus, it is difficult a priori to estimate how the shift of the nodes depends on wavenumber.

Figure~\ref{shiftvsredshift} shows the results of calculating the percentage of shift  for the four nodes at $k=0.057,0.088,0.115,0.14 \kvecMpc$ (vertical lines in Fig.~\ref{NodeShift}) as a function of redshift. Note that in this figure we use the transfer function for the linear spectrum evaluated at redshift $z=0$ (unlike the other figures where it was taken at $z=49$ as for the N-body simulation). The shift as a function of redshift for the node corresponding to the fundamental oscillation (node 1) approximately matches the shift in the peak of the correlation function due to mode-coupling, as discussed in the next section. This should not be surprising, the fundamental mode determines the location of the peak. 

The shift in the higher-order nodes (2 to 4) depends on node number, suggesting that mode-coupling does not cause a uniform (independent of $k$) overall shift. Note also that very similar shifts can be derived from our N-body results due to the close agreement of $B_{\rm MC}$ with RPT seen in the right panel of Fig.~\ref{NodeShift}.

These results seem at first glance in disagreement with those in~\cite{Durham2007}, who using N-body simulations with total volume very similar to ours concluded from the dark matter real space power spectrum that  there were no detectable shifts. However, many differences in the methods used make a direct comparison difficult, e.g. they use the rebinning method of~\cite{Petal07} to define smooth spectra, and they do a global fit looking for a uniform shift of $B$ with respect to $B_0$ (going to higher wavenumbers, up to $k=0.4\kvecMpc$) and simultaneously fitting for the damping due to the propagator. A global fit depends on the behavior of $B_{\rm MC}$ also far from the linear nodes, where its effect is smaller, however, is precisely there where the details of the damping model, through $g(k)$ in Eq.~(\ref{Bratio}), become important. We now try to perform a closer analysis to~\cite{Durham2007}.
Their implementation can be written as $B(k)=[B_0(k\alpha)-1]g(k)+1$, which in view of Eq.~(\ref{Bratio}) leads to~\footnote{It corresponds to choosing their damping $W=g$. They model $W$ as a Gaussian, in detail the propagator $G$ that enters into the definition of $g$ is better approximated by a Gaussian.},

\beq
B_0(k\alpha) = B_0(k)+\Big( \frac{1-g(k)}{g(k)}\Big) \, [B_{\rm MC}(k)-1],
\eeq
where $\alpha$ is a scaling parameter that characterizes an overall shift in $B$ with respect to $B_0$. Again, if $B_{\rm MC}$ were equal to unity there would be no overall shift, i.e. $\alpha=1$. If we use RPT to find the best fit $\alpha$ we obtain,

\beq
1-\alpha=0.5\%,0.35\%,0.2\%
\eeq
at $z=0,0.5,1$ respectively, where we have used scales up to $k=0.2 \kvecMpc$. 

One source of concern in the method of~\cite{Petal07} is that rebinning can introduce noise into the smooth spectrum that could wash out the oscillations in $B_{\rm MC}$. We have recalculated $B_{\rm MC}$ from the N-body simulations for a smooth power spectrum obtained by rebinning the nonlinear wiggly power $k^3P(k)$ within bins of width $\Delta k=0.044 \kvecMpc$ starting at our fundamental mode $2\pi/1280\approx 0.006 \kvecMpc$. The result is shown as triangles in  Fig.~\ref{NodeShift}, where we have suppressed error bars for clarity. We see that rebinning does introduce noise, although the overall features match what we found using our ``strictly smooth" method. However, changing the binning can affect results. To quantify this, following~\cite{Durham2007} we look for $\alpha$ by minimizing

\beq
\sum_i \Big(B_0(k_i\alpha) - B_0(k_i)-\Big( \frac{1-g(k_i)}{g(k_i)}\Big) \, [B_{\rm MC}(k_i)-1]\Big)^2 w_i^2,
\eeq
where the sum is over power spectrum bins up to $k=0.2 \kvecMpc$ and $w_i \equiv g(k_i) P(k_i)/\sigma_P(k_i)$. For the binning defined above ($\Delta k=0.044 \kvecMpc$, triangles in  Fig.~\ref{NodeShift}) we find $1-\alpha=0.75\%\pm0.2\%$ at $z=0$, consistent with our RPT results. If we bin in $P(k)$ instead of $k^3P(k)$ we get $1-\alpha=0.65\%\pm0.2\%$. However, trying different ways of rebinning (e.g. $\Delta k=0.04-0.06 \kvecMpc$) we found values of $1-\alpha$ that range from $1\%$ to $0\%$, which span many standard deviations, suggesting that the method of~\cite{Petal07} to define smooth spectra is not reliable enough to be used to detect shifts at below the percent level. These different rebinnings lead to fluctuations in the smooth reference spectrum of order $1\%$, while the power spectrum error bars for our total simulation volume are of order $0.5\%$ and $0.1\%$ at $k=0.05\kvecMpc$ and $0.2\kvecMpc$, respectively.

\begin{figure*}[ht!]
\begin{center}
\includegraphics[width=0.497\textwidth]{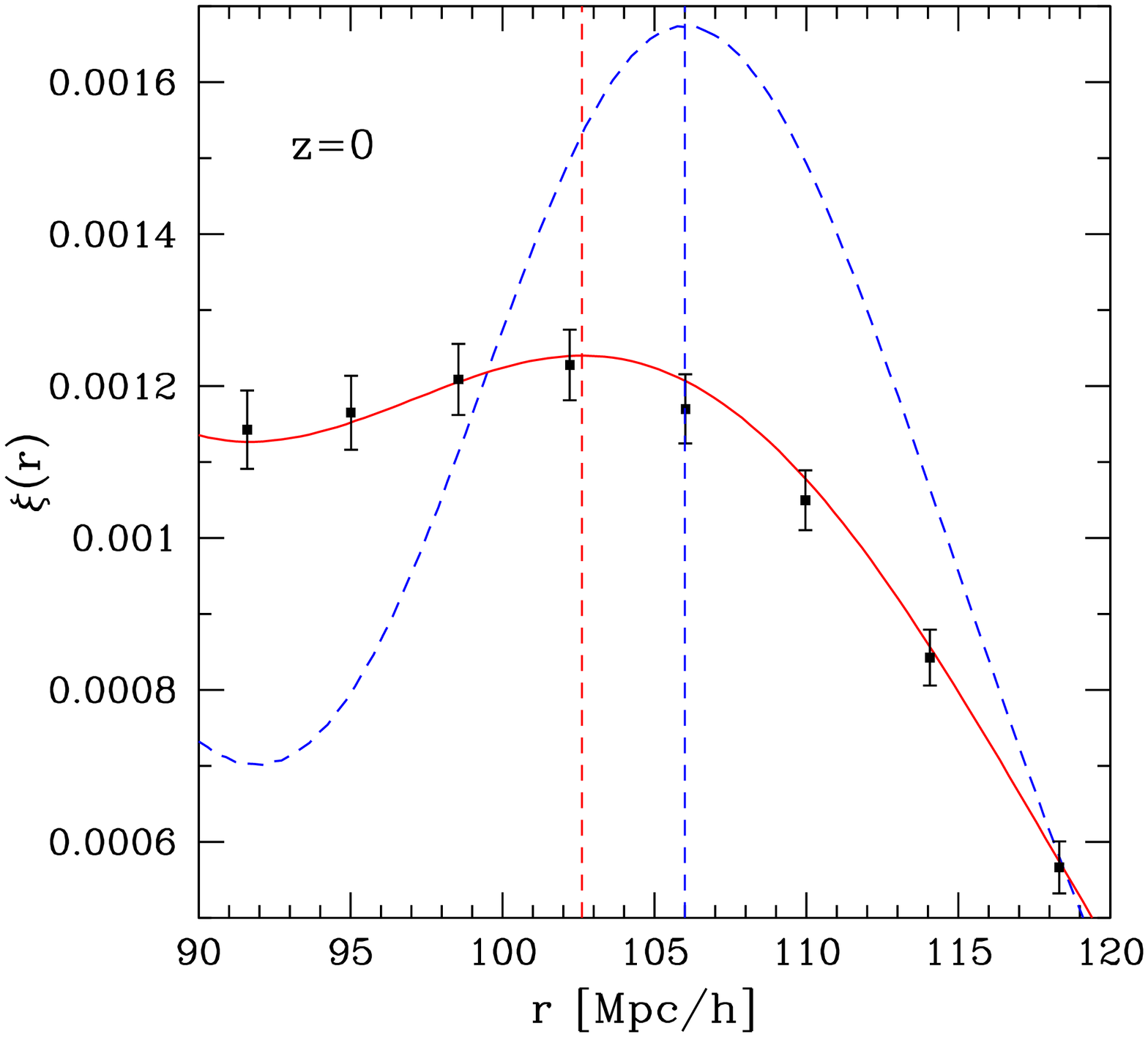} 
\includegraphics[width=0.497\textwidth]{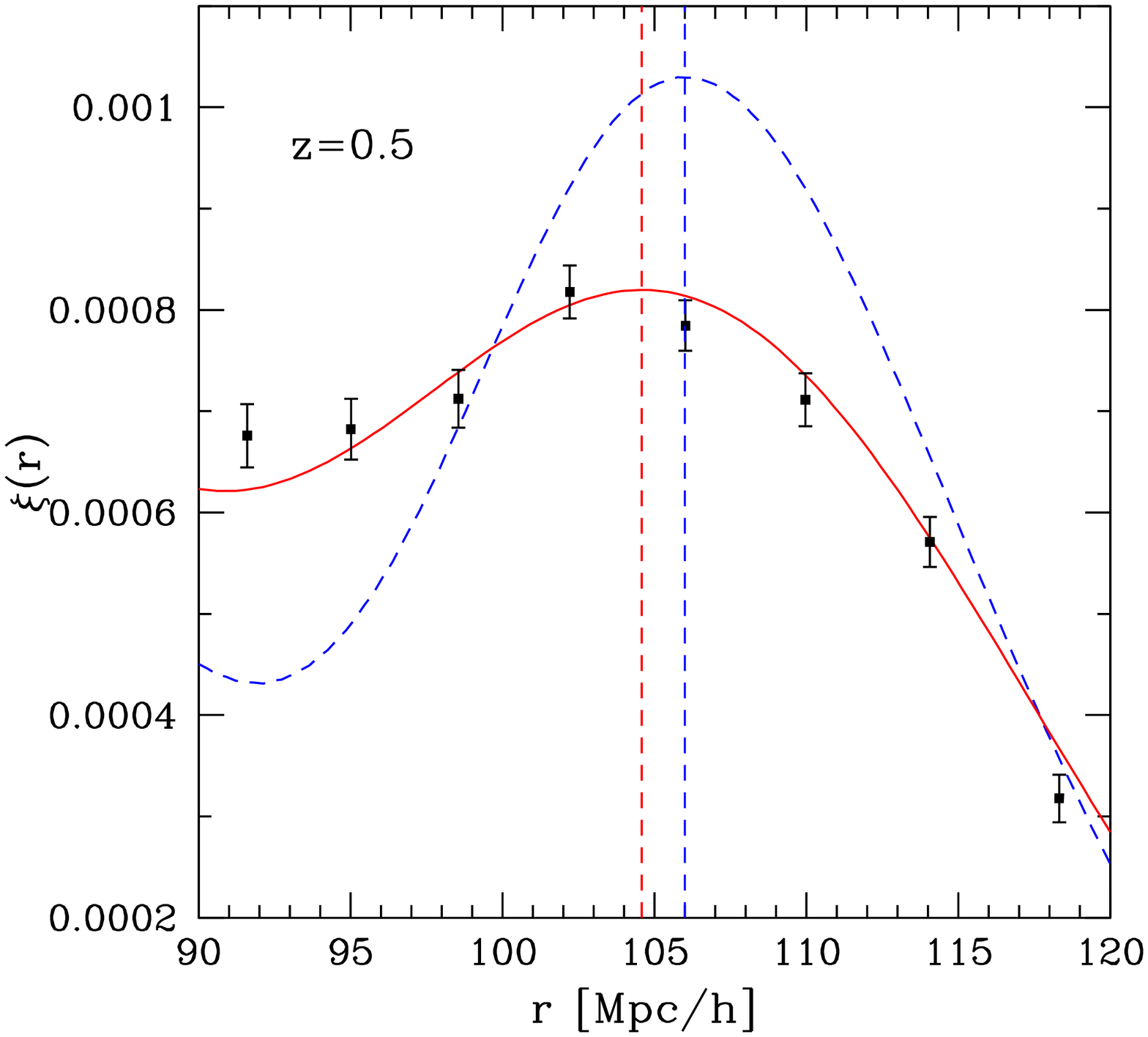}
\caption{A more detailed view of Figure~\ref{xinl} close to the acoustic peak. It shows the shift in the position of the acoustic peak of the correlation function according to RPT for $z=0$ (left panel) and $z=0.5$ (right panel). The linear peak (dashed line) is smeared out after being convolved with the propagator $G$, broadening it and shifting its position. The addition of the mode-coupling contribution $\xi_{\rm MC}$ provides another slight shift towards smaller scales, see Fig.~\protect\ref{pkand2PCF}. The sum of these two terms gives the prediction of RPT (solid line). Vertical dashed lines denote the local maxima of the linear and nonlinear two-point correlation function peaks. }
\label{zoomshift}
\end{center}
\end{figure*}

\subsection{Acoustic Peak in the TwoPoint Function}
\label{APshift}

Figure~\ref{zoomshift} provides a detailed view close to the acoustic peak in the two-point function, showing the linear theory (dashed line) and nonlinear (solid) correlation functions, together with vertical lines denoting the local peak maxima. The peak shifts towards small scales, and there are two sources of shifts. In addition to the smoothing of the acoustic peak, convolution by $G^2$ generates a shift in the acoustic peak towards smaller scales, because the linear acoustic peak is not quite symmetric about its maximum. This shift is referred to as ``apparent" in~\cite{RScube07}, and is typically taken into account in models used to analyze data, e.g.~\cite{ESW06,Durham2007}. In addition to this, there is another source of shifts, termed ``physical" in~\cite{RScube07}, which results from the mode-coupling contributions describing coherent infall of pairs in the model of~\cite{RScube07}. 

In Figure~\ref{shiftvsredshift} we separate the contribution to the peak shift in these two components, providing the percentual shift as a function of redshift due to mode-coupling alone, calculated in the one-loop approximation. Note that in this figure we have calculated the corresponding shifts when the transfer function is taken at $z=0$ rather than $z=49$ (as was used in the N-body simulations). This makes the linear peak sharper and thus more robust against nonlinearities, decreasing the shift due convolution with the propagator by roughly a factor of two, but only decreasing the mode-coupling shift by about fifteen percent. 

The mode-coupling shift decreases at high redshift, as expected.  Note that this calculation should be regarded as a lower bound to the shifts expected in galaxy redshift surveys, which include in addition the effects of redshift space distortions and galaxy bias, both of which increase their magnitude. See~\cite{RScube07} for calculations of the acoustic peak shift that include the effects of bias~\footnote{However, note that their analytic calculation, as well as their simulations (same as in this work), takes the transfer function at $z=49$.}.

Our results are in broad agreement with the estimates of the acoustic peak shift made in~\cite{GuBeSm2006} using {\ttfamily halofit} to describe the nonlinear power spectrum. They found a shift of $2.4\%$ at $z=0$, comparable to ours, e.g. if we include the shift due to the propagator at $z=0$ we obtain 1.9\%. Our calculations are also in good agreement with those in~\cite{RScube07} for $z\ga0.5$, at lower redshifts their method overestimates the shifts. As they note, this is due to the breakdown of standard perturbation theory. Note that, unlike RPT, none of these methods are able to separate the total shift into its two different components. 

Our results for the shifts generated by mode-coupling are also expected given the results from simulations presented in~\cite{RScube07} for halos. In that case a fit for the overall shift was attempted in which only the convolution (``apparent") shift was allowed for. The best fit model had statistically significant residuals that made apparent that a second source of shifts due to mode-coupling was present. Furthermore, note that since this was found for {\em point-like halos}, the physics responsible for shifts must be unrelated to virialized halo profiles. In the next section we explain in detail how these shifts arise.

Finally, note that the shift in the peak of the two-point function agrees with that found for the fundamental oscillation node in the power spectrum at all redshifts, as expected from the fact that the fundamental oscillation sets the scale for the peak of the two-point function. We show in the next section that in fact both methods (correlation function peak and power spectrum oscillation nodes) are subject to the same systematics, in the sense that if we ignore oscillations in the mode-coupling spectrum the mode-coupling shift in the two-point function also vanishes.

\begin{figure*}[ht!]
\begin{center}
\includegraphics[width=0.49\textwidth]{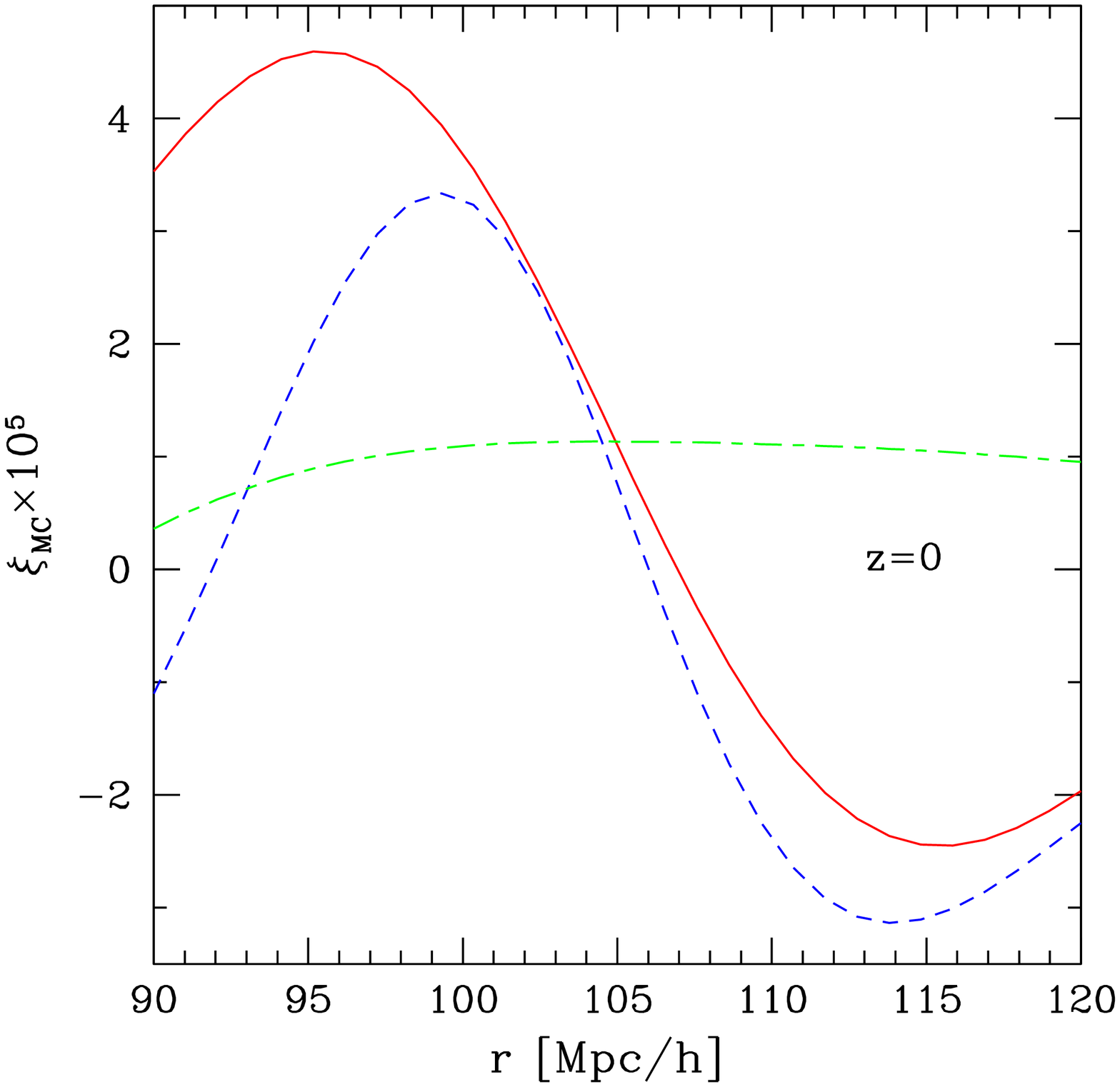}
\includegraphics[width=0.49\textwidth]{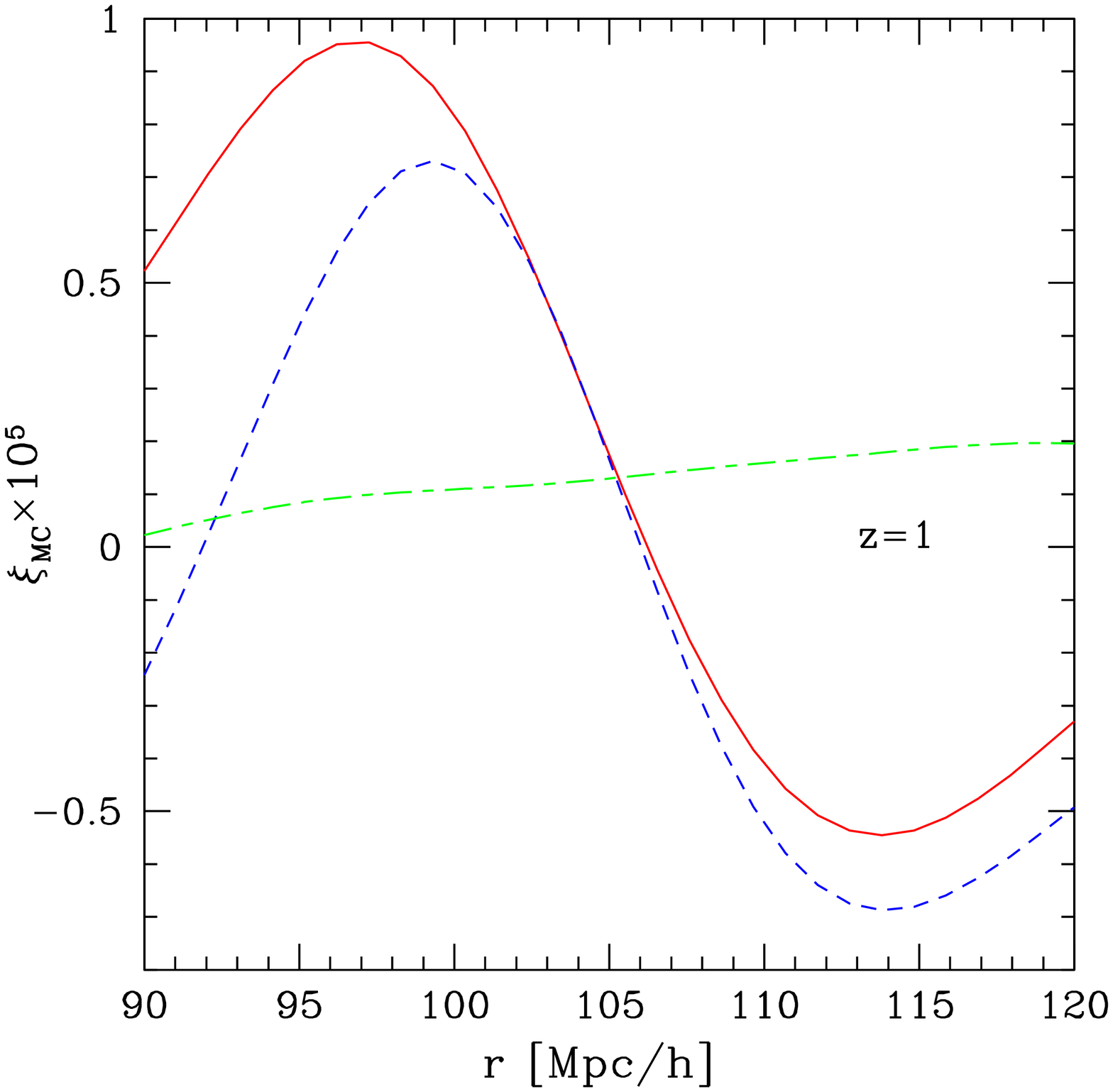} 
\caption{The mode-coupling two-point correlation function as predicted by RPT (solid lines) at redshift $z=0$ (left) and $z=1$ (right). The dashed lines show a fit to it of the form given by the second term in Eq.~(\protect\ref{phenomodel}), where we only allow for an arbitrary constant prefactor. The short dash-long dash line denotes the mode-coupling correlation function corresponding to a smooth initial power spectrum and therefore has no particular feature at the acoustic scale. Thus, correcting for smooth broadband tilts in the power spectrum does not correct for the shift of the two-point function peak.}
\label{modelMC}
\end{center}
\end{figure*}

\subsection{Why shifts?}
\label{WS}

The shift of the acoustic peak generated by $\xi_{\rm MC}$ is simple to understand from first principles. To do so, standard PT (which gives the low-$k$ behavior of $P_{\rm MC}$) suffices here~\footnote{For another way of seeing how shifts arise naturally, see~\cite{RScube07}. The $\v\cdot\nabla$ terms in the equations of motion we highlight here lead to the flow of characteristics and the erasure of the acoustic peak in the divergence of the pairwise infall velocities $\Theta$ discussed in~\cite{RScube07}.}. In this case we have

\beq
P_{\rm MC}(k) \simeq 2 \int [F_2(\k-\q,\q)]^2 P_{\rm lin}(|\k-\q|) P_{\rm lin}(q)\, d^3q,
\label{PmcLowk}
\eeq
where $F_2$ denotes the second-order PT kernel ($\tvk=\k/k$),
\beqa
F_2(\k,\q) &=& \frac{17}{21} +\frac{1}{2}\, \tvk \cdot \tvq \, \Big(\frac{k}{q}+\frac{q}{k}\Big) 
+ \frac{4}{21}\, \Big(\tvk_i \tvk_j -\frac{1}{3}\de_{ij}\Big) \nonumber \\
&&\times \Big(\tvq_i \tvq_j -\frac{1}{3}\de_{ij}\Big)
\label{F2k}
\eeqa

The physical interpretation of these terms is as follows: the first term (monopole) describes the growth of perturbations in the spherical collapse model to second order; the second term (dipole) represents the transport of matter by the velocity field encoded by the $\v\cdot \nabla$ terms in the equations of motion; the last term (quadrupole) describes the impact of tidal gravitational fields in the growth of structure. The excellent agreement between  RPT and simulations in Fig.~\ref{modecouplingpower} at BAO scales tells us that Eq.~(\ref{PmcLowk}) is quite accurate for the discussion that follows.

The important contribution for our purposes is the second term in Eq.~(\ref{F2k}), which is the only one that leads to derivatives of the linear correlation function. Squaring the $F_2$ kernel, the lowest order term that leads to a shift comes from the double product of the first and second terms, which upon Fourier transformation leads to a contribution of the kind,
\beq
 \xi'_{\rm lin}(r)\ \xi^{(1)}_{\rm lin}(r),
\label{xishift}
\eeq
where 
\beq
\xi^{(1)}_{\rm lin}(r) \equiv \hat{r}\cdot \nabla^{-1}\xi_{\rm lin}(r) = \int {d^3k \over k}\, P_{\rm lin}(k)\, j_1(kr)
\eeq
is an integrated version of the linear two-point function and thus very flat about the acoustic peak. On the other hand, the derivative involved in Eq.~(\ref{xishift}) leads to a net shift towards small scales, since $\xi'_{\rm lin}$ is positive (negative) to the left (right) of the peak. Close to the peak this can be thought as inducing a tilt in the linear correlation function. Note that shifts appear as a consequence of gravitational instability being nonlocal, and by the same physics that generates a non-zero bispectrum at the largest scales. The spherical model where kernels such as $F_2$ are replaced by their spherical average does not lead to shifts of the acoustic peak. Another example that does not lead to shifts is the lognormal model~\cite{RScube07}.

It's important to note that the two sources of shifts discussed above operate in a different way, that due to $G^2$ is multiplicative (or convolution) while that due to mode coupling is additive. This has interesting consequences. The shift generated by $G^2$ will be absent if the linear peak were symmetric, or sharp enough compared to the width of $G^2$. On the other hand, how much shift mode coupling generates depends on how much smoothing $G$ has done: since its effect is additive it will be able to modify more strongly a weaker peak. Note that these considerations apply also in Fourier space. Furthermore, the smoothing scale involved in $G$ and the mode-coupling power are not very sensitive to features in the linear power spectrum. These considerations may help explain why in~\cite{Ma2007} placing a large peak (by BAO standards) in the initial power spectrum did not seem to generate shifts.

Finally, note that Eq.~(\ref{PmcLowk}) also explains the main features of $P_{\rm MC}$ seen in Fig.~\ref{NodeShift}. First, the convolution of wiggly spectra leads to oscillations by beating the oscillation at $|\k-\q|$ with that at $q$; note this also leads to a result that oscillates not about the mode-coupling power of smooth linear spectra, but rather about the average of the product of wiggly linear powers, thus explaining  the drift seen in the solid lines in the left panel of Fig.~\ref{NodeShift}. Another way of seeing this, is to calculate Eq.~(\ref{PmcLowk}) for a power-law power spectrum of arbitrary spectral index $n$, as done in~\cite{SF96}. The result is that $P_{\rm MC}$ is a strong decreasing function of $n$: nonlinear growth is enhanced for a more negative spectral index. Again, the main cause for this effect are the $\v\cdot\nabla$  terms that lead to Eq.~(\ref{xishift}); this is also in fact the reason why the skewness of a smoothed density field increases as $n$ become more negative. This dependence on the local spectral index means that for a wiggly linear spectrum, the nonlinear growth will be enhanced (suppressed) with respect to the smooth case on the nodes of oscillations that have negative (positive) derivative, creating the out of phase correction seen in Fig.~\ref{NodeShift}. This, in turn, leads to the node shifts derived in Eq.~(\ref{NLnodes}).

Since the shift in the peak of the two-point function and the power spectrum oscillation nodes have the same  physical origin, it should not be surprising that switching off one of them also makes the other go away. We now show that making the mode-coupling power smooth leads to no mode-coupling shift for the peak of the correlation function. Figure~\ref{modelMC} shows (short dashed-long dashed line) the Fourier transform of the smooth mode-coupling power spectrum obtained by evolving under RPT the smooth linear spectrum obtained as discussed in section~\ref{PSW}. We see that the resulting mode-coupling two-point function $\xi_{\rm MC}^{\rm s}$ is flat about the acoustic scale, as expected since the smooth initial conditions have no knowledge of a preferred scale, as opposed to $\xi_{\rm MC}$ for wiggly initial conditions (solid line) which is positive (negative) to the left (right) of the linear acoustic scale. For scales $r \la 80 \Mpc$ or $r \ga 130 \Mpc$, $\xi_{\rm MC}^{\rm s}$ approximately follows $\xi_{\rm MC}$, at levels consistent with the drift seen in $B_{\rm MC}$ about unity in the left panel of Fig.~\ref{NodeShift}. 

From this discussion we see that if the mode-coupling power were smooth ($B_{\rm MC}=1$), there would no shift in the peak of the two-point function caused by mode-coupling. In other words, {\em correcting for smooth broadband tilts in the power spectrum does not correct for the shift of the two-point function peak.}

\subsection{A Simple Model}

We can now estimate how the shifts generated by mode coupling depend on galaxy bias and redshift distortions. Indeed, in the presence of local bias perturbation theory tells us that Eq.~(\ref{PmcLowk}) gets modified into~\cite{Heavensetal1998,RScube06,McDonald06}
\beqa
P_{\rm MC}^{\rm gal}(k) &=& 
2b_1b_2 \int F_2(\k-\q,\q) P_{\rm lin}(|\k-\q|) P_{\rm lin}(q)\, d^3q \nonumber \\ 
&+& b_1^2\ P_{\rm MC}(k) + \frac{b_2^2}{2} \int P_{\rm lin}(|\k-\q|) P_{\rm lin}(q)\, d^3q,
\nonumber \\ &&
\label{PmcLowkgal}
\eeqa
where $b_1$ and $b_2$ are the linear and quadratic bias parameters. We see that an additional term that induces shift in the acoustic peak of the two point function arises from nonlinear bias, proportional to $b_1b_2$. It has the same form as Eq.~(\ref{xishift}) and will lead to enhanced shifts compared to the linear bias case for galaxies that preferentially populate high-mass halos, in which case $b_2>0$, in agreement with the arguments given in~\cite{RScube06} for the power spectrum. In addition, redshift-space distortions amplify the effect of the second term in the $F_2$ kernel responsible for shifts by $(1+\beta \mu^2)$~\cite{ZBisp}, where $\beta = f/b_1$, $f=d\ln\Dp/d\ln a$, and $\mu$ is the cosine of the wave-vector along the line of sight, under the approximation that the mapping from real to redshift space can be treated perturbatively.

Motivated by the above discussion we can suggest a simple phenomenological model to take into account nonlinear effects close to the acoustic peak that may be useful in practical applications. One can try to fit the observed galaxy correlation function monopole in redshift space by 

\beq
\xi_{\rm obs}(r) =  A\, [{\rm e}^{-r^2/\sigma^2} \otimes \xi_{\rm lin}](r) + B\,  \xi'_{\rm lin}(r)
\label{phenomodel}
\eeq
where we have used that $\xi^{(1)}_{\rm lin}$ in Eq.~(\ref{xishift}) varies very little around the acoustic peak, and $A$, $B$ and $\sigma$ are constants that can be fit for. From linear bias and redshift distortions, one expects $A \sim b_1^2 (1+2\beta/3+\beta^2/5)$, $B \sim A\, \xi^{(1)}_{\rm lin}(r_p) \ll A$ and $\sigma^2\sim 4\sigma_v^2(1+f/3)$, where $r_p$ is the position of the acoustic peak.  Similarly, in the method that uses the power spectrum oscillation nodes, one may want to add a small correction that has an out of phase oscillation, which is analogous to the second term in Eq.~(\ref{phenomodel}). Finally, note that these correction terms coming from mode coupling are intrinsically second order in the power spectrum or correlation function, thus one may want to include this information in the modeling.

It is beyond the scope of this paper to study such models in detail, but for the same reason that a result from first principles such as Eq.~(\ref{Ghighk}) was later shown to be useful phenomenologically in the presence of bias and redshift distortions (see e.g.~\cite{ESW06,Huff07}) we would like to stress that the second term in Eq.~(\ref{phenomodel}), despite being derived here invoking a number of approximations, is likely to prove similarly useful. To see this, we show in Fig.~\ref{modelMC} the full result for $\xi_{\rm MC}$ in RPT and a fit to it of the form given by the second term in Eq.~(\protect\ref{phenomodel}), where we only allow for an arbitrary constant prefactor. We see that despite the simplicity of the model it captures the behavior rather well. One could certainly improve the fit by including more terms that result from Eq.~(\ref{PmcLowk}). 

In addition, the relevance of Eq.~(\ref{phenomodel}) for more complicated situations where bias and redshift distortions are present can be appreciated from looking at the form of the residuals to fits of halo-halo correlation functions in redshift space~(Fig.~5 in~\cite{RScube07}) that includes only the first term in Eq.~(\ref{phenomodel}). These residuals have a form very well matched by the second term in Eq.~(\ref{phenomodel}), shown in Fig.~\ref{modelMC} by dashed lines.

\section{Conclusions}

We calculated the nonlinear power spectrum and two-point correlation function in renormalized perturbation theory (\cite{paper1,paper2}, RPT), with particular emphasis on the description of baryon acoustic oscillations (BAO). Our predictions are in excellent agreement with detailed measurements in numerical simulations, as shown in Figs.~\ref{nlacoustic} through~\ref{zoomshift}.

In this paper we have concentrated on a single cosmological model at different redshifts; this was driven by the desire to compare our predictions to simulations at the percent level at BAO scales, which requires very large volumes, e.g. our total simulation volume is  $\simeq105(\Gpc)^3$. Calculating RPT predictions for different cosmological models is rather simple, while having corresponding simulations with the required precision at BAO scales is significantly time consuming, this will be considered in a forthcoming paper~\cite{paper4}. 

RPT describes the power spectrum (or correlation function) as the sum of two contributions, Eq.~(\ref{equationP}), both of which are sensitive to nonlinear effects. One term dominates at  large scales and is proportional (via the propagator $G$) to the initial power spectrum at the same scale, reducing to linear theory in the large-scale limit. We already tested in~\cite{paper2} the RPT prediction for the propagator against numerical simulations, for both densities and velocities as a function of redshift. The propagator is responsible for the damping of BAO in the power spectrum, or smoothing of the acoustic peak in the two-point function. 

The second contribution, the mode-coupling power spectrum $P_{\rm MC}$, describes the generation of power at a given scale due to the coupling of two or more modes from different scales. Calculation of this in RPT was the main focus of this paper (see Appendix~\ref{RPTpower} for technical details). Figure~\ref{modecouplingpower} shows that the mode-coupling power in the simulations grows as dictated by RPT. Mode coupling modifies the acoustic pattern in the power spectrum, inducing oscillations out of phase with those in linear theory,  and this generates a shift towards small scales in the acoustic peak of the two-point function. Our predictions for these effects are also in excellent agreement with simulations.

The halo model describes two-point statistics in a way that resembles RPT. However, there are important differences that we highlighted in this paper.  Although the sum of the one-halo and two-halo terms as written in section~\ref{RPTvsH} describes the power spectrum in simulations to within $\simeq 15\%$, this is highly misleading because  the two-halo term is too large and the one-halo term too small in the range of scales between linear and virialized regimes, $0.05 \kvecMpc \la k \la 1 \kvecMpc$. This is simply because in this range the physics that operates is neither linear theory nor due to virialized halos. This is obvious when one tries to use the halo model to describe higher-order statistics such as the bispectrum: using linear theory plus virialized halos fails miserably, and one must include weakly nonlinear physics~\cite{SSHJ01}. Once this is recognized, it is simply not optional to include or not such contributions for two-point statistics. Similarly, as emphasized in~\cite{RScube06}, one must pay attention to nonlinearities in the large-scale bias relation when describing two-point statistics of halos or galaxies at large scales.

Fixing the two-halo term by adding mode-coupling contributions and correct damping of the linear spectrum is possible by incorporating the missing physics (e.g. by RPT); however, it is less clear at present how to avoid the wrong low-$k$ limit of the one-halo term (see Fig.~\ref{modecouplingpower}). In any case, the test any improved model must pass is very simple: it must match the cross-correlation of initial and final density fields and the mode-coupling power, instead of only using the power spectrum or correlation function as a diagnostic.

Having demonstrated that the mode-coupling power predicted by RPT is in very good agreement with simulations, we discussed what are the implications for the evolution of the BAO. We showed that the mode-coupling power contains oscillations that are essentially out of phase with those in the linear spectrum. This leads to shifts towards smaller scales in the position of the oscillation nodes used in the BAO analysis method proposed in~\cite{Petal07}.  We showed that this mode-coupling contribution, when Fourier transformed, naturally leads to a shift of the acoustic peak in the two-point function towards smaller scales, through terms proportional to the derivative of the linear correlation function, Eq.~(\ref{xishift}).  
We also showed that correcting for smooth broadband tilts in the power spectrum does not correct for the shift of the two-point function peak.

We gave predictions for how these mode-coupling shifts depend on redshift, which should be considered as a robust lower limit to the more realistic case when galaxy bias and redshift distortions are also present. See~\cite{RScube07} for calculations that also include the effects of galaxy bias on the acoustic peak shift.

Once the source of shifts is identified, however, it is easy to see what to expect when redshift-space distortions and galaxy bias are included. This understanding allows us to suggest a simple physically motivated model that can be used in dealing with such effects in observations. In fact, the model naturally predicts the kind of pattern observed in~\cite{RScube07} as residuals when the halo correlation function in redshift space is fit to a model that only includes the effects of smoothing of the peak. 

Finally, we would like to emphasize that the nonlinearities we focus on in this paper do operate at the acoustic scale. 
The same mode-coupling effects that we discussed here are responsible for generating a non-zero bispectrum at large scales; e.g. this is the reason why the skewness in spheres of diameter equal to the size of the acoustic scale is about forty percent at $z=0$~\cite{CosmoBisp}. Most importantly, the detailed form of these couplings, e.g. Eq.~(\ref{F2k}), has been observed in galaxy surveys, giving rise to the dependence of the bispectrum on triangle shape~\cite{PSCzB}. It might be possible, in fact, to empirically determine corrections for clustering systematics in two-point statistics from the measured three-point function.

\acknowledgments

We thank R.~Angulo, C.~Baugh, F.~Bernardeau, P.~Fosalba, G.~Gabadadze, E.~Gazta\~naga, A.~Gruzinov, A.~Hamilton, D.~Hogg, L.~Hui, N.~Padmanabhan, S.~Pueblas, E.~Sefusatti, M.~Tegmark and M.~White for useful discussions. R.~S.~thanks the Aspen Center for Physics for hospitality while some of this work was done. Special thanks to R.~Sheth and R.~Smith for comments on a preliminary version of the manuscript and many useful discussions. MC acknowledges support from Spanish Ministerio de Ciencia y Tecnologia (MEC), project AYA2006-06341 with EC-FEDER funding and
research project 2005SGR00728 from  Generalitat de Catalunya. This work was partially supported by NSF AST-0607747 and NASA NNG06GH21G.

\appendix
\section{Calculating the  power spectrum in RPT}
\label{RPTpower}

The nonlinear evolution of the power spectrum in the framework of RPT is made out of two basic ingredients: the nonlinear propagator $G$ and the mode coupling power $P_{\rm MC}$ \cite{paper1}. Recalling Eq.~(\ref{equationP}) we have for the density field power spectrum,
\beq
P(k,z)=G^2(k,z) \, P_0(k) + P_{\rm MC}(k,z),
\label{eqP}
\eeq
where $P_0(k)$ is the initial spectrum of fluctuations. In \cite{paper2} we computed the propagator $G$ while in this paper we completed the description by computing $P_{MC}$ at the scales relevant for BAO. Although a detailed analysis of these calculations will be given elsewhere~\cite{paper4} we present 
in this appendix the expression for the contributions to $P_{\rm MC}$ that we use throughout this paper (e.g. Eq.~(\ref{PmcLoops})), together with the approximations involved.

In \cite{paper1} we showed that $P_{\rm MC}$ can be written as an infinite sum of contributions expressed in terms of only three resummed quantities, the nonlinear propagator $G_{ab}(k,z)$, the full vertex $\Gamma_{abc}(\k,\k_1,\k_2)$ and the nonlinear power spectrum $P_{ab}(k,z)$, where sub-indices here each take on two values since they refer to either density contrast or velocity divergence fields. To lowest order these quantities reduce to the {\em linear} propagator ($\eta \equiv \ln \Dp$), 
\beq
g_{ab}(z) = \frac{{\rm e}^\eta}{5}
\Bigg[ \begin{array}{rr} 3 & 2 \\ 3 & 2 \end{array} \Bigg] -
\frac{{\rm e}^{-3\eta/2}}{5}
\Bigg[ \begin{array}{rr} -2 & 2 \\ 3 & -3 \end{array} \Bigg],
\label{linearprop}
\eeq
which one can think of as four growth factors, i.e. how much does the density or velocity grow with respect to the initial densities and velocities (thus making a two by two matrix) and, similarly, four decaying modes; the symmetrized {\em vertex}, 
\beqa
\gamma_{121}^{({\rm s})}(\k,\k_1,\k_2)&=&\delta_{\rm D}(\k-\k_{12})\ {(\k_{12} \cdot \k_1)\over{ 2\ k_1^2}},  \label{linearvertex1}\\
\gamma_{112}^{({\rm s})}(\k,\k_2,\k_1)&=&\delta_{\rm D}(\k-\k_{12})\ {(\k_{12} \cdot \k_2)\over{ 2\ k_2^2}},  \label{linearvertex2}\\
\gamma_{222}^{({\rm s})}(\k,\k_1,\k_2)&=&\delta_{\rm D}(\k-\k_{12})\ {|\k_{12}|^2 (\k_1 \cdot \k_2 )\over{2\ k_1^2 \ k_2^2}},
\label{linearvertex}
\eeqa
with $\k_{12}\equiv \k_1+\k_2$ and $\gamma_{abc}=0$ otherwise; and the {\em linear} power spectra 
\beq
P^{\rm lin}_{ab}(k,z) = [D_+(z)]^2\, P_0(k)\,
\Bigg[ \begin{array}{rr} 1 & 1 \\ 1 & 1 \end{array} \Bigg]
\label{linearPmatrix}
 \eeq
from standard PT \cite{PTreview,Sco00,paper1}. Note that we are assuming growing mode initial conditions where the initial density contrast $\delta$ and velocity divergence $\nabla\cdot \u$ (where $\v=-f {\cal H} \u$, with $f=d\ln\Dp/d\ln a$, ${\cal H} =d\ln a/d\tau$ and $\tau$ conformal time) are the same random field. This is the reason why all the linear spectra in Eq.~(\ref{linearPmatrix}) are determined by a single spectrum, $P_0$. For the same reason, the combination $\tilde G_a \equiv G_{ab}\cdot (1,1)$ represents the propagator for density and velocities for growing mode initial conditions, and it appears frequently. For instance, the density propagator in Eq.~(\ref{eqP}) is actually given by $G\equiv \tilde G_1=G_{11}+G_{12}$.

As we mentioned in Sec.~\ref{nlps} each contribution to $P_{\rm MC}$ is dominant only in a narrow range of scales centered at an increasingly higher value of $k$. We found that considering the first two contributions to $P_{\rm MC}$ was sufficient to describe the range of scales relevant to BAO (see Figs.~\ref{PDet} and~\ref{modecouplingpower}). Since in diagrammatic language they correspond to a one-loop diagram  and a two-loop diagram \cite{paper1} we refer to them as $P_{\rm MC}^{\rm 1loop}$ and $P_{\rm MC}^{\rm 2loop}$, respectively.

In \cite{paper2} we studied in detail the nonlinear propagator $G_{ab}(k,z)$ and derived an analytical prescription for it, valid for all times and scales, that showed a remarkably good agreement with measurements in numerical simulations well into the nonlinear regime (see Sec.~\ref{nlps} for a brief account of the main ideas and procedure behind this derivation). Using better simulations, we have found that our prediction was slightly overestimating the new measurements by a few percent at intermediate scales. Out of the few reasons that could cause this slight mismatch the leading possibility  is a small contribution from subdominant diagrams in the large-$k$ limit resummation for $G_{ab}$. Research under way will confirm whether this is the reason or not, for this paper we proceed to make a small correction to our calculated propagator in~\cite{paper2} under this assumption. It is important to note that once this is done, the prediction for the nonlinear power spectrum is fully determined. As we discussed in connection with Fig.~\ref{NodeShift}, the simulations used in this work have a propagator that decays slightly {\em slower} than our modified RPT prescription; that is, our procedure is not a fit to the propagator measured in the simulations.

As mentioned in Sec.~\ref{nlps} the prediction for $G_{ab}$ was obtained by matching its low-$k$ behavior, computable exactly using one-loop PT, with its large-$k$ limit, where it can be shown to decay exponentially as $\exp(-k^2 \sigma_v^2/2)$ with $\sigma_v^2=\frac{1}{3}\int P_{\rm L}(k,z) d^3q/q^2$ (see Eq.~(\ref{Ghighk})). In turn, this limit was obtained by resumming an infinite sub-set of diagrams in the series expansion for the propagator. In \cite{paper1} we argued that this sub-set gives the dominant contribution to the resummed propagator. Although this remains true we have estimated the contribution coming from sub-leading diagrams in the resummation of the large-$k$ limit and found that it can reproduce the few percent level disagreement with the simulations at all redshifts.

The way we estimated this subdominant contribution is as follows. We notice that a sub-set of these subleading diagrams leads to power spectrum resummation. In this case the nonlinear spectrum, instead of the linear, should be used to compute the nonlinear propagator in its large-$k$ limit (i.e. $\sigma_v^2 \rightarrow \alpha^2(z) \sigma_v^2$, with $\alpha^2(z)$ given by $\int P_{\rm nl}(k,z) d^3q/q^2/\int P_{\rm lin}(k,z) d^3q/q^2$). For simplicity, we computed this factor using {\ttfamily halofit} to describe $P_{\rm nl}(k,z)$, and found that it grows monotically from $1$ at high redshift to $\alpha \sim 1.05$ at $z=0$. Thus we  multiply by $\alpha^2(z)$ the exponents in each component of $G_{ab}$ in Eq. (41) of~\cite{paper2}. 

For the present work, we use the functional form for $G_{ab}$ given in Eqs. (41) of \cite{paper2}, corrected as described above, to compute  $P_{\rm MC}^{\rm 1loop}$ and $P_{\rm MC}^{\rm 2loop}$. In addition, the full vertex is set at its lowest order, i.e. the symmetrized vertex is given by Eqs.~(\ref{linearvertex1}-\ref{linearvertex}), while the nonlinear spectrum that enters into the calculation of $P_{\rm MC}$ is given by $P_{ab}(k,z)={\tilde G}_a(k,z)  P_0(k) {\tilde G}_b(k,z)$, which in a iterative scheme  based on Eq.~(\ref{eqP}) would correspond to the first step. 
Under this approximation the one-loop contribution to $P_{\rm MC}$ is given by (see Eq. (33) in \cite{paper1}),
\begin{widetext}
\beq
P_{\rm 1\,Loop}(\k,\eta)= 2  \int d^3q \,\Lambda_{\rm 1\,Loop} (\k,\q,\eta) \ \Lambda_{\rm 1\,Loop} (-\k,-\q,\eta) \, P_0(q) P_0(|\k-\q|), 
\label{apponeloop}
\eeq
with
\beq
\Lambda_{\rm 1\,Loop} (\k,\q,\eta)=\int_0^\eta ds_1 \,G_{1b}(|\k|,\eta,s_1)\,\gamma^{(s)}_{bcd}(\k,{\bf q},\k-{\bf q})\,\tilde G_c(\q,s_1,0) \tilde G_d(|\k-\q|,s_1,0),
\label{lambdaoneloop}
\eeq
where the time variable $\eta=\ln\Dp$, and $\tilde G_a = G_{a1}+G_{a2}$. The second contribution, corresponding to a two-loop diagram, is given by (see Eq. (34) in \cite{paper1}),
\beq
P_{\rm 2\,Loop}(\k,\eta)=16 \! \int \! d^3q \int \! d^3p \,\Lambda_{\rm 2\,Loop} (\k,\q,{\bf p},\eta) \ \Lambda_{\rm 2\,Loop} (-\k,-{\bf p},-\q,\eta) \, P_0(q)P_0(p)P_0(|\k-{\bf p}-\q|), 
\label{apptwoloop}
\eeq
with,
\beqa
\Lambda_{\rm 2\,Loop}(\k,\q,{\bf p},\eta)=\int_0^\eta \!\! ds_1 \int_0^{s_1} \!\! ds_2 \,  G_{1b}(|\k|,\eta,s_1)\, \gamma_{bcd}^{(s)}(\k,\k-\q,\q)\, G_{cf}(|\k-\q|,s_1,s_2)\, \tilde G_d(\q,s_1,0) \nonumber \\
\gamma_{fgh}^{(s)}(\k-\q,{\bf p},\k-{\bf p}-\q) \, \tilde G_g (|{\bf p}|,s_2,0) \, \tilde G_h (|\k-{\bf p}-\q|,s_2,0). \qquad \qquad \ \
\label{lambdatwoloop}
\eeqa
\end{widetext}

These lowest order assumptions for the full vertex and nonlinear power in the computation of $P_{\rm MC}$ are expected to be very robust at low-$k$, e.g. if one were to stop at the one-loop level where the dominant behavior is dictated by the decay of the propagator. At the two-loop level one should also check the impact on $P_{\rm MC}^{\rm 1loop}$ coming from the renormalization of the vertex or a second step in the iterative scheme. At present we are studying these effects and the results will be presented in \cite{paper4} and \cite{paper5}. 

Nonetheless it can be concluded a posteriori that this  lowest order approach for $\Gamma$ and $P$ works very well at the BAO scales given the excellent agreement shown in Figs.~\ref{nlacoustic}-\ref{modecouplingpower}. However, 
we expect that at the one-percent level, some of our approximations within the framework of  RPT are likely to begin to fail. We can briefly summarize the main issues at play,

\begin{enumerate}

\item[i)] The matching ansatz for the propagator done in~\cite{paper2}, and the size of sub-leading diagrams in the large-$k$ limit. 

\item[ii)]  The so-called ``vertex renormalization"~\cite{paper1} and the inclusion of mode coupling power in the r.h.s of Eqs.~(\ref{apponeloop}, \ref{apptwoloop}) must be quantified. 

\item[iii)] The incorrect cosmological dependence of decaying modes is expected to play a role in the nonlinear regime, as perturbations do not remain in the growing mode after many interactions. This enters through our assumption that the linear decaying mode obeys $D_{-}\propto \Dp^{-3/2}$, see Eq.~(\ref{linearprop}). This is only true as long as we can set $f/\Omega_m^2 \simeq \Omega_m^{1/9} $ to unity~\cite{paper1}.

\item[iv)] The impact of shell crossing, which so far we have neglected since we work in the fluid approximation.

\end{enumerate}

The first set of issues  were effectively corrected by renormalizing $\sigma_v$ as described above. But further research is needed to incorporate this, if needed, in a stricter way. The three remaining effects should have negligible impact on the modeling of BAO since they are expected to become important at smaller scales, but they are approachable and will be discussed in future work. 

Regarding the last issue listed above, it was recently argued in~\cite{sticky} that the impact of shell-crossing on the power spectrum at BAO scales is of order one percent, and that any agreement at the percent level (such as that shown in Fig.~\ref{PDet}) must be ``accidental" unless justified by solving the collisionless Boltzmann equation. In principle percent-level corrections from shell crossing are possible, although we disagree that the calculation in~\cite{sticky} addresses the issue. The impact of shell crossing must be estimated from the difference between the collisionless solution and the fluid approximation. In~\cite{sticky}, on the other hand, the argument is based on a comparison between two extremely different situations (and none of them is the fluid limit): the collisionless case versus the ``sticky" dark matter scenario where collisions have a huge cross section. It is not clear that the difference between these two situations would be a good estimate of the effect of shell crossing (i.e. the fluid limit may be a much better approximation in the collisionless case).
Most importantly, the halo model as written in section~\ref{RPTvsH} was used in place of the collisionless solution, and the halo model with delta function profiles (but same mass function) was assumed to describe the collisional solution. As discussed in section~\ref{RPTvsH}, the one-halo term is about a factor of four too small at $k\simeq 0.1 \kvecMpc$ (see Fig.~\ref{modecouplingpower}), and consequently the two-halo term (where most of the effect quoted in~\cite{sticky} is coming from) is too large by about the same factor.
Given all these assumptions, we do not feel the conclusions in~\cite{sticky} are warranted. 

One can estimate directly from Fig.~\ref{modecouplingpower} when the contribution from virialized profiles (where shell crossing has definitely taken place) becomes comparable to the mode-coupling power in our simulations by extrapolating the one-halo term to small scales, and this happens at $k\simeq 1 \kvecMpc$.


\begin{thebibliography}{10}

\bibitem{BAO} D.~J.~Eisenstein, W.~Hu, M.~Tegmark, Astrophys. J. {\bf 504}, L57 (1998); A.~Cooray, W.~Hu, D.~Huterer, M.~Joffre, Astrophys. J. {\bf 557}, L7 (2001); W.~Hu, Z.~Haiman, Phys. Rev. D {\bf 68}, 063004 (2003); C.~Blake, K.~Glazebrook, Astrophys. J. {\bf 594}, 665 (2003); H.-J.~Seo, D.~J.~Eisenstein, Astrophys. J. {\bf 598}, 720 (2003); C.~Blake, K.~Glazebrook, Astrophys. J. {\bf 631}, 1 (2005); H.~Zhan, L.~Knox, Astrophys. J. {\bf 644}, 663 (2006)

\bibitem{wiggles} P.~J.~E.~Peebles, J.~T..~Yu, Astrophys. J. {\bf 162}, 815 (1970); R.~Sunyaev, Ya.-B. Zel'dovich, Astrophys. J. Suppl. Ser. {\bf 7}, 3 (1970)

\bibitem{BAOdetect} D.~J.~Eisenstein et al., Astrophys. J. {\bf 633}, 560 (2005)

\bibitem{BAOobs} S.~Cole et al., Mon. Not. R. Astron. Soc. {\bf 362}, 505 (2005); M.~Tegmark et al., Phys. Rev. D {\bf 74}, 123507 (2006); H\"utsi, Astron. Astrophys. {\bf 449}, 891 (2006); Percival et al., Astrophys. J. {\bf 657}, 51 (2007); N.~Padmanabhan et al., astro-ph/0605302 (2006)

\bibitem{Durham2007} R.~Angulo, C. M. Baugh, C. S. Frenk, C. G. Lacey, astro-ph/0702543 (2007).

\bibitem{Nbody} A.~Meiksin, M.~White, J.~A.~Peacock, Mon. Not. R. Astron. Soc. {\bf 304}, 851 (1999); H.-J.~Seo, D.~J.~Eisenstein, Astrophys. J. {\bf 598}, 720 (2003); M.~White, Astropart. Phys. {\bf 24}, 334 (2005); V.~Springel et al., Nature {\bf 435}, 629 (2005)

\bibitem{Huff07} E.~Huff, A.~E.~Schultz, M.~White, D.~J.~Schlegel, M.~S.~Warren, Astroparticle Phys. {\bf 26}, 351 (2007)

\bibitem{ESSS06}D.~J.~Eisenstein, H.-J.~Seo, E.~Sirko, D.~N.~S.~Spergel, arXiv:astro-ph/0604362

\bibitem{Ma2007}Z.~Ma, astro-ph/0610213, (2007)

\bibitem{SW06} A.~E.~Schultz, M.~White, Astroparticle Phys. {\bf 25}, 172 (2006)

\bibitem{komatsu} D. Jeong, E. Komatsu, 2006, astro-ph/0604075

\bibitem{GuBeSm2006} 
J.~Guzik, G.~Bernstein, R.~E.~Smith, astro-ph/0605594 (2006).

\bibitem{ESW06} D. J. Eisenstein, H.-J. Seo, M. White, astro-ph/064361 (2006)

\bibitem{RScube06}R.~E.~Smith, R.~Scoccimarro, R.~K.~Sheth, Phys. Rev. D {\bf 75}, 063512 (2007).

\bibitem{MaPie07} S.~Matarrese, M,~Pietroni, astro-ph/0702653; S.~Matarrese, M,~Pietroni, astro-ph/0703563

\bibitem{RScube07}R.~E.~Smith, R.~Scoccimarro, R.~K.~Sheth, astro-ph/0703620 (2007)

\bibitem{Halofit} R. E. Smith, et al, Mon. Not. R. Astron. Soc. {\bf 341}, 1311 (2003)

\bibitem{paper1} M. Crocce and R. Scoccimarro, Phys. Rev. D {\bf 73}, 063519 (2005)

\bibitem{paper2} M. Crocce and R. Scoccimarro, Phys. Rev. D {\bf 73}, 063520 (2005)

\bibitem{Valageas06} P.~Valageas,  astro-ph/0611849 (2006)

\bibitem{McDonald07} P.~McDonald, Phys. Rev. D {\bf 75}, 043514 (2007) 

\bibitem{B96} S. Bharadwaj, Astrophys. J. {\bf 472}, 1 (1996)

\bibitem{paper4} M. Crocce and R. Scoccimarro, in preparation (2007).

\bibitem{PTreview} F. Bernardeau, S. Colombi, E. Gazta\~naga, R. Scoccimarro, Phys. Rep. {\bf 367}, 1 (2002)

\bibitem{transients1} R. Scoccimarro, Mon. Not. R. Astron. Soc. {\bf 299}, 1097 (1998)

\bibitem{transients2} M. Crocce, S. Pueblas, R. Scoccimarro, Mon. Not. R. Astron. Soc. {\bf 373}, 369 (2006)

\bibitem{gadget2} V. Springel, MNRAS {\bf 364}, 1105 (2005)

\bibitem{bbks} J. M. Bardeen, J. R. Bond, N. Kaiser and A. S. Szalay, Astrophys. J. {\bf 304}, 15 (1986) 

\bibitem{paper5} F. Bernardeau, M. Crocce, R. Scoccimarro, in preparation (2007).

\bibitem{loopPower}
N.~Makino, M.~Sasaki, Y.~Suto, Phys. Rev. D {\bf 46}, 585 (1992); B.~Jain, E.~Bertschinger, Astrophys. J. {\bf 431}, 495 (1994); C.~M.~Baugh, G.~Efstathiou, Mon. Not. R. Astron. Soc. {\bf 270}, 183 (1994);
E.~L.~{\L}okas, R.~Juszkiewicz, F.~R.~Bouchet, E.~Hivon, Astrophys. J. {\bf 467}, 1 (1996);
R.~Scoccimarro, J.~A.~Frieman, Astrophys. J. {\bf 473}, 620 (1996)

\bibitem{LA85} K.~Heitmann, P.~M.~Ricker, M.~S.~Warren, S.~Habib, Astrophys. J. Suppl. {\bf 160}, 28 (2005)

\bibitem{haloreview}A.~Cooray, R.~K.~Sheth, Phys. Rep. {\bf 372}, 1 (2002)

\bibitem{Yang03}X.~Yang, H.~J.~Mo, F.~C.~van den Bosch, Mon. Not. R. Astron. Soc. {\bf 339}, 1057 (2003)

\bibitem{Zheng04}Z.~Zheng, Astrophys. J. {\bf 610}, 61 (2004)

\bibitem{Tinker05}J.~L.~Tinker, D.~H.~Weinberg, Z.~Zheng, I.~Zehavi, Astrophys. J. {\bf 631}, 41 (2005)

\bibitem{McDonald06}P.~ McDonald, Phys. Rev. D {\bf 74}, 103512 (2006)

\bibitem{ST99}R.~K.~Sheth, G.~Tormen, Mon. Not. R. Astron. Soc. {\bf 308}, 119 (1999).

\bibitem{Zel65}Ya.~B.~Zel'dovich, Adv. Astron. Astrophys. {\bf 3}, 241 (1965)

\bibitem{Peebles80}P. J. E. Peebles, The Large-Scale Structure of the Universe, \S 28, 
           Princeton University Press, Princeton, NJ, 1980.

\bibitem{GGRW86} M.~H.~Goroff, B.~Grinstein, S.-J. Rey, M.~B.~Wise, Astrophys. J. {\bf 311}, 6 (1986)

\bibitem{Netal07} T.~Nishimichi, H.~Ohmuro, M.~Nakamichi, A.~Taruya, K.~Yahata, A.~Shirata, S.~Saito, H.~Nomura, K.~Yamamoto, Y.~Suto, astro-ph/07051589 (2007)

\bibitem{Petal07}W.~J.~Percival, S.~Cole, D.~J.~Eisenstein, R.~C.Nichol, J.~A.~Peacock, A.~C.~Pope, A.~S.~Szalay, 
astro-ph/07053323 (2007)

\bibitem{BG03}C.~Blake, K.~Glazebrook, Astrophys. J. {\bf 594}, 665 (2003)

\bibitem{SF96}R.~Scoccimarro, J.~A.~Frieman, Astrophys. J. {\bf 473}, 620 (1996)

\bibitem{Heavensetal1998}A.~F.~Heavens, S.~Matarrese, L.~Verde, Mon. Not. R. Astron. Soc. {\bf 301}, 797 (1998).

\bibitem{ZBisp} E.~Hivon, F.~R.~Bouchet, S.~Colombi, R.~Juszkiewicz, Astron. Astrophys. {\bf 298}, 643 (1995); 
L.~Verde, A.~F.~Heavens, S.~Matarrese, L.~Moscardini, Mon. Not. R. Astron. Soc. {\bf 300}, 747 (1998);  R.~Scoccimarro, H.~M.~P.~Couchman, J.~A.~Frieman,  Astrophys. J. {\bf 517}, 531 (1999)

\bibitem{SSHJ01} R.~Scoccimarro, R.~K.~Sheth, L.~Hui, B.~Jain, Astrophys. J. {\bf 546}, 20 (2001).

\bibitem{CosmoBisp} E. Sefusatti, M. Crocce, S. Pueblas, R. Scoccimarro, Phys. Rev. D, {\bf 74}, 023522 (2006).

\bibitem{PSCzB} H.A. Feldman, J.A. Frieman, J.N. Fry, R. Scoccimarro, Phys. Rev. Lett {\bf 86}, 1434  (2001). 

\bibitem{Sco00}R.~Scoccimarro, Annals New York Academy Sciences {\bf 927}, 13 (2001).

\bibitem{sticky} N.~Afshordi, Phys. Rev. D {\bf 75}, 021302(R) (2007)

\end{thebibliography}
\end{document}